\documentclass[useAMS,usenatbib,usegraphicx]{mn2e}

\usepackage{amssymb}
\usepackage{amsmath}
\usepackage{booktabs}
\PassOptionsToPackage{hyphens}{url}\usepackage{hyperref}
\usepackage{multirow}
\usepackage{times}
\usepackage[hyphenbreaks]{breakurl}

\title[Multiwavelength variability of Ark\,120]
  {A deep X-ray view of the bare AGN Ark\,120.  III.  X-ray timing analysis and multiwavelength variability}
\author[A.P. Lobban]
  {A.P.~Lobban,$^1$\thanks{e-mail: \href{mailto:a.p.lobban1@keele.ac.uk}{a.p.lobban1@keele.ac.uk}}
  D.~Porquet$^{2,3}$, J.N.~Reeves$^1$, A.~Markowitz$^{4,5}$, E.~Nardini$^6$, N.~Grosso$^{2,3}$
  \\
  $^1$Astrophysics Group, School of Physical and Geographical Sciences, Keele University, Keele, Staffordshire, ST5 5BG, U.K. \\
  $^2$Universit\'{e} de Strasbourg, Observatoire astronomique de Strasbourg, CNRS, UMR 7550, 11 rue de l'Universit\'{e}, F-67000 Strasbourg, France \\
  $^3$Aix Marseille Univ, CNRS, LAM, Laboratoire d'Astrophysique de Marseille, Marseille, France \\
  $^4$Nicholas Copernicus Astronomical Center, ul. Bartycka 18,
00-716 Warsaw, Poland \\
  $^5$Center for Astrophysics and Space Sciences, Univ. of California, San Diego, MC 0424, La Jolla, California, 92093-0424, U.S.A. \\
  $^6$INAF - Osservatorio Astrofisico di Arcetri, Largo E. Fermi 5, I-50125 Firenze, Italy \\
  }
\date{\today; accepted for publication in MNRAS}

\pagerange{\pageref{firstpage}--\pageref{lastpage}} \pubyear{2017}

\def\LaTeX{L\kern-.36em\raise.3ex\hbox{a}\kern-.15em
    T\kern-.1667em\lower.7ex\hbox{E}\kern-.125emX}

\begin{document}

\label{firstpage}

\maketitle

\begin{abstract}

We present the spectral/timing properties of the bare Seyfert galaxy Ark\,120 through a deep $\sim$420\,ks {\it XMM-Newton} campaign plus recent {\it NuSTAR} observations and a $\sim$6-month {\it Swift} monitoring campaign.  We investigate the spectral decomposition through fractional rms, covariance and difference spectra, finding the mid- to long-timescale ($\sim$day--year) variability to be dominated by a relatively smooth, steep component, peaking in the soft X-ray band.  Additionally, we find evidence for variable Fe\,K emission red-ward of the Fe\,K$\alpha$ core on long timescales, consistent with previous findings.  We detect a clearly-defined power spectrum which we model with a power law with a slope of $\alpha \sim 1.9$.  By extending the power spectrum to lower frequencies through the inclusion of {\it Swift} and {\it RXTE} data, we find tentative evidence of a high-frequency break, consistent with existing scaling relations.  We also explore frequency-dependent Fourier time lags, detecting a negative (`soft') lag for the first time in this source with the 0.3--1\,keV band lagging behind the 1--4\,keV band with a time delay, $\tau$, of $\sim 900$\,s.  Finally, we analyze the variability in the optical and UV bands using the Optical/UV Monitor on-board {\it XMM-Newton} and the UVOT on-board {\it Swift} and search for time-dependent correlations between the optical/UV/X-ray bands.  We find tentative evidence for the U-band emission lagging behind the X-rays with a time delay of $\tau = 2.4 \pm 1.8$\,days, which we discuss in the context of disc reprocessing.

\end{abstract}


\section{Introduction} \label{sec:introduction}

Active galactic nuclei (AGN) are routinely observed to emit strongly across the entire electromagnetic spectrum with the broad-band spectral energy distribution comprising both thermal and non-thermal emission components \citep{Shang11}.  Normally peaking at ultraviolet (UV) wavelengths, the dominant energy output of Seyfert galaxies is generally considered to arise from thermal emission from material in the inner parts of a geometrically-thin, optically-thick accretion disc surrounding the supermassive black hole (SMBH; \citealt{ShakuraSunyaev73}).  The location of the UV-emitting material is typically 10--1\,000\,$r_{\rm g}$\footnote{The gravitational radius is defined as $r_{\rm g} = GM_{\rm BH}/c^{2}$.} from the central black hole, depending on the properties of the accretion flow.  This thermal UV emission is then thought to be responsible for producing the X-ray continuum, which is commonly power-law in shape, via inverse-Compton scattering of soft thermal photons via an optically-thin `corona' of hot ($T \sim 10^{9}$\,K) electrons, usually within a few tens of $r_{\rm g}$ from the black hole \citep{HaardtMaraschi93}.

Given the compact nature of AGN and the great distances at which we observe them, their central regions cannot be directly resolved with the current generation of observatories.  However, information about the geometry, structure and physical processes that dominate these systems can be indirectly inferred via alternative methods.  In particular, variability studies have proven to be powerful techniques for increasing our understanding of the innermost regions of AGN.  One such technique for analysing variability is ``reverberation mapping'' \citep{BlandfordMcKee82}, which has proven successful in helping to unravel the structure of the broad-line region (BLR) in many AGN (e.g. \citealt{Clavel91}; \citealt{Peterson91}).  While reverberation mapping is an excellent tool for mapping the BLR, applying similar techniques can, in principle, be used to map the structure of the accretion disc.  In many AGN, strongly variable UV emission is observed (e.g. \citealt{Collin01}).  While the UV emission does not vary as rapidly as it does in the X-ray band \citep{MushotzkyDonePounds93}, one might expect to observe a correlation between the variability if the local accretion rate is responsible for modulating the X-ray and UV emission regions.  In this scenario, there are two favoured mechanisms to couple the variability - (i) UV photons being Compton up-scattered to X-rays by a corona of hot electrons (e.g. \citealt{HaardtMaraschi91}), and (ii) thermal reprocessing of X-rays in the disc (\citealt{GuilbertRees88,Collier98,CackettHorneWinkler07}).  In either case, it is the time taken for the light to cross between the two emission sites that governs the observed time delays.

To better understand how the various emission mechanisms are connected, it is important to analyse correlations between the emission in separate wavelength bands.  To date, correlations between the X-ray and UV bands have been observed in numerous AGN (e.g. \citealt{Nandra98}, \citealt{Cameron12}).  Additionally, correlations between the X-ray and optical bands have also been observed, but on much longer timescales.  These are often discussed in the context of disc reprocessing (e.g. \citealt{Arevalo08}; \citealt{Breedt09}; \citealt{Arevalo09}; \citealt{Breedt10}; \citealt{AlstonVaughanUttley13}; \citealt{McHardy14}; \citealt{Edelson15}).

In terms of the X-ray variability, a number of techniques are available to help probe the variability, often in a variety of model-independent ways.  Such methods include analysis of the rms spectrum and the covariance spectrum.  Recent developments have also been made in order measure the frequency-dependence of the observed variability through such techniques as estimating the PSD (power spectral density) and `X-ray time lags' (i.e. time delays between different X-ray energy bands).  These are commonly seen to exhibit similar behaviour across a variety of sources but with the measured amplitudes and frequencies roughly scaling inversely with the mass of the black hole (e.g. \citealt{Lawrence87}; \citealt{UttleyMcHardyPapadakis02}; \citealt{Vaughan03}; \citealt{Markowitz03}; \citealt{Papadakis04}; \citealt{McHardy04}; \citealt{McHardy06}; \citealt{Arevalo08}).  In addition, the amplitude of the X-ray rms variability is routinely seen to scale with the source flux in a linear way.  This is known as the `rms-flux relation'.  Again, this appears to be a common X-ray variability property that holds over a large timescale range and is present in both AGN and XRBs (e.g. \citealt{UttleyMcHardy01}; \citealt{Vaughan03}; \citealt{Gaskell04}; \citealt{UttleyMcHardyVaughan05}).

Ark\,120 is a nearby ($z = 0.0327$; \citealt{OsterbrockPhillips77}) Seyfert galaxy at a distance of 142\,Mpc (assuming standard cosmological parameters; i.e. $H_{\rm 0} = 71$\,km\,s$^{-1}$; $\Omega_{\Lambda} = 0.73$; $\Omega_{\rm M} = 0.27$).  It is the X-ray-brightest of the known subclass of `bare' AGN ($F_{\rm 0.3-10\,keV} \sim 7 \times 10^{-11}$\,erg\,cm$^{-2}$\,s$^{-1}$) where, through analysis of high-resolution grating data, \citet{Vaughan04} constrained the column density of any line-of-sight X-ray warm absorber to be at least a factor of $10$ lower than a typical type-1 Seyfert galaxy.  As such, Ark\,120 offers the clearest view of the central regions close to the black hole and is an ideal candidate for studying the intrinsic high-energy process associated with AGN --- e.g. Comptonization, reflection, etc.

The X-ray spectrum of Ark\,120 generally shows the standard features expected from a type-1 Seyfert with a dominant hard power law, a smooth soft excess emerging $< 2$\,keV \citep{Brandt93}, a complex Fe\,K emission profile and associated Compton reflection hump emerging $> 10$\,keV (\citealt{Nardini11}).  The source is additionally shown to be variable at optical-to-X-ray wavelengths (e.g. \citealt{Gliozzi17}) with a pronounced drop in broad-band flux in 2013 \citep{Matt14}.  Reverberation mapping studies show that the mass of the black hole in Ark\,120 is high: $M_{\rm BH} = 1.5 \pm 0.2 \times 10^{8}$\,$M_{\odot}$ \citep{Peterson04}.

Here we report on the variability properties of Ark\,120, concentrating on data obtained from a deep {\it XMM-Newton} campaign in 2014 (AO-12 Large Programme; PI: D. Porquet), but also including {\it Swift} and {\it NuSTAR} data.  This is the third in a series of papers based on this campaign.  In Reeves et al. (2016; hereafter: paper I) we analyzed the high-resolution grating spectra with {\it XMM-Newton} and {\it Chandra} and, for the first time in this source, discovered a series of ionized soft X-ray emission lines, which we interpret as arising from an X-ray component of the optical-UV sub-pc BLR, suggestive of significant columns of X-ray emitting gas outside of our line-of-sight.  In Nardini et al. (2016; hereafter: paper II) we analyzed the Fe\,K properties of the source finding excess emission both red-ward and blue-ward of the strong near-neutral Fe\,K$\alpha$ emission line at $\sim$6.4\,keV.  The wings of the line are variable on different timescales and may arise from transient hotspots on the surface of the accretion  a few tens of $r_{\rm g}$ from the black hole.  A fourth paper presents a detailed analysis of the broad-band spectrum (\citealt{Porquet17}; hereafter: paper IV).  Here, we focus on the spectral-timing properties of the source with an emphasis on broad-band variability.

\section{Observations and data reduction} \label{sec:data_reduction}

\subsection{{\it XMM-Newton}} \label{sec:xmm_observations}

Ark\,120 was observed with {\it XMM-Newton}  \citep{Jansen01} on 2003-08-24 (duration $\sim$112\,ks), 2013-02-18 ($\sim$130\,ks) and four times in 2014 over four consecutive orbits, each with typical durations of $\sim$130\,ks: 2014-03-18, 2014-03-20, 2014-03-22 and 2014-03-24 (hereafter, referred to as 2014a, 2014b, 2014c and 2014d, respectively).  Here, we focus on data acquired with the European Photon Imaging Camera (EPIC)-pn and the co-aligned Optical/UV Monitor (OM; \citealt{Mason01}), which provides simultaneous optical/UV coverage.  The Reflection Grating Spectrometer (RGS; \citealt{denHerder01}) data are presented in paper I.  All raw data were processed, following standard procedures, using version 15.0 of the {\it XMM-Newton} Scientific Analysis Software (\textsc{sas}\footnote{\url{http://xmm.esac.esa.int/sas/}}) package, with the 2017 release of the Current Calibration File.

\subsubsection{The EPIC-pn} \label{sec:epic-pn}

The EPIC-pn camera was operated in small-window mode (SW; $\sim$71\,per cent `live time') using the thin filter.  As Ark\,120 is a bright source, to alleviate pile-up concerns, we do not utilize data from the EPIC Metal-Oxide Semiconductor (MOS) cameras and we only consider single EPIC-pn events (i.e {\tt PATTERN} $= 0$) $> 0.3$\,keV.  In order to minimize the contribution from the background, source events were extracted from circular regions of 20\,arcsec radius, which were centred on the source\footnote{Source events from a circular region of 40\,arcsec radius [corresponding to a larger encircled energy fraction; e.g. $\sim$92\,per cent instead of $\sim$88\,per cent at $\sim$1.5\,keV (values taken from the {\it XMM-Newton} user handbook)]were also extracted but no significant difference was found in the results, other than a small increase in the count rate.}.  We extracted background events from larger regions ($\sim$50\,arcsec) away from the source and the edges of the small-window CCD.  For most of the {\it XMM-Newton} observations, the background level was relatively stable.  However, the data quality was further optimised by removing time periods of background flaring from the subsequent analysis (typically just a few ks per observation).  Such events usually occur due to proton flaring towards the end of each observation, before the pericentre passage.  We note that the 0.3--10\,keV background count rate is low  ($< 1$\,per cent of the source rate) for all six observations.  The observed broad-band EPIC-pn count rates, exposure times and fluxes are detailed in Table~\ref{tab:xmm_log}.

\begin{table}
\centering
\begin{tabular}{l c c c c}
\toprule
& \multirow{2}{*}{{\it XMM}} & Exposure \\
Date & \multirow{2}{*}{Rev.} & [Duration] & Count Rate & Flux \\
& & (ks) & (ct\,s$^{-1}$) & (erg\,cm$^{-2}$\,s$^{-1}$) \\
\midrule
2003-08-24 & 0679 & 71 [112] & 12.3 [18.1] & $7.4 \times 10^{-11}$ \\
2013-02-18 & 2417 & 80 [130] & 5.6 [8.4] & $4.0 \times 10^{-11}$ \\
2014-03-18 & 2614 & 83 [133] & 12.5 [18.5] & $7.6 \times 10^{-11}$ \\
2014-03-20 & 2615 & 84 [132] & 10.6 [15.6] & $6.6 \times 10^{-11}$ \\
2014-03-22 & 2616 & 84 [133] & 11.6 [16.1] & $7.3 \times 10^{-11}$ \\
2014-03-24 & 2617 & 85 [133] & 10.7 [15.7] & $6.7 \times 10^{-11}$ \\
\bottomrule
\end{tabular}
\caption{Observation log of the {\it XMM-Newton} EPIC-pn observations.  `Rev.' refers to the {\it XMM-Newton} revolution/orbit number.  Exposure times refer to the net time (i.e. `live time') of the instrument after background filtering, while the numbers in parentheses relate to the total duration of the observation.  The background-subtracted count rates (derived from single pixel events only) and fluxes are quoted from 0.3--10\,keV.  The count rates in parentheses are those obtained from a 40\,arcsec source radius.  Note that the count rates are the `intrinsic' observed rates in the sense that they do not fold in the loss of counts due to the large EPIC-pn SW mode deadtime.}
\label{tab:xmm_log}
\end{table}

\subsubsection{The Optical/UV Monitor} \label{sec:om}

The OM consists of six primary filters: V (peak effective wavelength $=$ 5\,430\,\AA), B (4\,500\,\AA), U (3\,440\,\AA), UVW1 (2\,910\,\AA), UVM2 (2\,310\,\AA) and UVW2 (2\,120\,\AA).  During each of the observations in 2014, we acquired $\sim$5 $\sim$1.2\,ks exposures in both ``imaging'' and ``fast'' mode cycling through the V, B, U, UVW1 and UVM2 filters before spending the remainder of the observation acquiring exposures with the UVW2 filter.  Each observation resulted in a total of $\sim$80 imaging exposures.  In the case of the 2013 observation, a series of snapshots were acquired only with the UVW1, UVM2 and UVW2 filters while in 2003, monitoring was only undertaken with the UVW2 filter.  See Table~\ref{tab:om_log} for an observation log.

\begin{table}
\centering
\begin{tabular}{l c c c c c c c c c c c c}
\toprule
\multirow{2}{*}{{\it XMM}} & \multicolumn{12}{c}{OM Filter} \\
\multirow{2}{*}{Obs.} & \multicolumn{2}{c}{V} &  \multicolumn{2}{c}{B} &  \multicolumn{2}{c}{U} & \multicolumn{2}{c}{UVW1} &  \multicolumn{2}{c}{UVM2} &  \multicolumn{2}{c}{UVW2} \\
& $\#$ & E & $\#$ & E & $\#$ & E & $\#$ & E & $\#$ & E & $\#$ & E \\
\midrule
2003 & 0 & 0 & 0 & 0 & 0 & 0 & 0 & 0 & 0 & 0 & 80 & 80 \\
2013 & 0 & 0 & 0 & 0 & 0 & 0 & 10 & 34 & 10 & 39 & 10 & 44 \\
2014a & 5 & 6 & 5 & 6 & 5 & 6 & 5 & 6 & 5 & 6 & 55 & 66 \\
2014b & 5 & 6 & 5 & 6 & 5 & 6 & 5 & 6 & 25 & 30 & 35 & 42 \\
2014c & 5 & 6 & 5 & 6 & 5 & 6 & 5 & 6 & 5 & 6 & 55 & 66 \\
2014d & 5 & 6 & 5 & 6 & 5 & 6 & 5 & 6 & 5 & 6 & 55 & 66 \\
\bottomrule
\end{tabular}
\caption{Observation log of the {\it XMM-Newton} OM observations showing the number of images per filter ($\#$) and the summed exposure time (E) in units of ks.}
\label{tab:om_log}
\end{table}

All imaging mode and fast mode data were processed using the \textsc{sas} tasks \textsc{omichain}\footnote{\url{http://xmm.esac.esa.int/sas/current/doc/omichain/}} and \textsc{omfchain}\footnote{\url{http://xmm.esac.esa.int/sas/current/doc/omfchain/}} ($\Delta t = 10$\,s), respectively, as part of {\it XMM-Newton}'s \textsc{sas}.  The routines take into account all necessary calibration processes (e.g. flat-fielding) and run source detection algorithms before performing aperture photometry on all detected sources.  Detector-dead-time and coincidence-loss corrections are then applied to the resultant count rates.

\subsection{{\it Swift}} \label{sec:swift_observations}

Ark\,120 was also observed with {\it Swift} \citep{Gehrels04} with a long-term $\sim$6-month monitoring campaign over the period 2014-09-04 - 2015-03-15 (obsID: 00091909XXX).  {\it Swift} performed a total of 86 observations, each generally separated by $\sim$2\,days and with typical durations of $\sim$1\,ks.  Here, we use data from {\it Swift}'s X-ray Telescope (XRT; \citealt{Burrows05}) and the co-aligned Ultra-Violet/Optical Telescope (UVOT; \citealt{Roming05}).

\subsubsection{The XRT} \label{sec:xrt}

As Ark\,120 is known to be a bright source, the XRT was operated in ``windowed timing mode'' (WT), which provides 1-dimensional imaging data at the orientation of the spacecraft roll angle.  We extracted useful XRT counts from 0.3--10\,keV using the online XRT products builder\footnote{\url{http://swift.ac.uk/user_objects/}} \citep{EvansBeardmorePage09}, which undertakes all required processing and provides background-subtracted spectra and light curves.  The XRT count rates include the systematic error arising from an inaccurate knowledge of the source position on the CCD when in WT mode\footnote{\url{http://www.swift.ac.uk/xrt_curves/docs.php\#systematics}}.  These systematics are understood and calibrated by the {\it Swift} team and are applied by the XRT products builder.  We also note that, since the WT count rate for Ark\,120 is significantly less than $\sim$100\,ct\,s$^{-1}$, the data should not be affected by pile-up (see \citealt{Romano06}).  During processing, two XRT observations were flagged as having unreliable count rates due to a WT-mode centroid not being found for the source\footnote{\url{http://www.swift.ac.uk/xrt_curves/docs.php\#wtnocent}}.  As such, we exclude the XRT data from these two observations [obsID: 00091909072 (2015-02-05); 00091909080 (2015-02-21)] from our subsequent analysis.  Over the entire ~$\sim$6-month period, the total XRT exposure was $\sim$85.6\,ks with time-averaged corrected count rate and flux of $\sim$1.6\,count\,s$^{-1}$ and $\sim$4.86 $\times 10^{-11}$\,erg\,cm$^{-2}$\,s$^{-1}$ from 0.3--10\,keV, respectively.

Due to observational constraints, {\it Swift} is not always able to observe the target source, occasionally resulting in an observation being split up into $> 1$ `snapshots' (i.e. where each snapshot has a fractional exposure $= 1$).  In the case of the Ark\,120 monitoring campaign, across the 84 useful XRT observations, this results in a total of 96 snapshots (fractional exposure $= 1$), 93 of which have exposure times $> 100$\,s.  Subsequently, an XRT `observation' refers to the entirety of a {\it Swift} pointing while an XRT `snapshot' refers to time periods where the fractional exposure $= 1$.

\subsubsection{The UVOT} \label{sec:uvot}

In addition to the XRT, we also utilize data acquired with the UVOT, which provides simultaneous UV/optical coverage with a possible wavelength range of $\sim$1\,700--6\,500\,\AA in a 17' $\times$ 17' field.  During each {\it Swift} pointing, data were acquired with either of the U or UVM2 filters (peak effective wavelengths: 3\,465 and 2\,246\,\AA, respectively), generally alternating between the two (i.e. pointings separated by $\sim$4\,days per filter).  Visually inspecting the UVOT images shows that the pointings were steady, with a total of 43 usable frames for each of the U and UVM2 filters.  All exposures were roughly $\sim$1\,ks in length (total U-band exposure: $\sim$39.5\,ks; total UVM2-band exposure: $\sim$41.9\,ks)\footnote{{\it Swift} also obtained two snapshots with the UVW1 ($\sim$2\,600\,\AA) filter and three snapshots with the UVW2 filter ($\sim$1\,928\,\AA), although these are not used in our correlation analysis here.}.  We extracted source counts from Ark\,120 using the HEA{\sc soft}\footnote{\url{http://heasarc.nasa.gov/lheasoft/}} (v.6.18) task \textsc{uvotsource}.  This undertakes aperture photometry while using the latest calibration database (CALDB) to make corrections - e.g. scaled background subtraction and coincidence loss.  The source extraction radius was 5\,arcsec while, for the background counts, a radius of 25\,arcsec was used, taken from a region of blank sky separate from the source.

\subsection{{\it NuSTAR}} \label{sec:nustar}

Ark\,120 has been observed by {\it NuSTAR} \citep{Harrison13} twice: 2013-02-18 (ID: 60001044002) and 2014-03-22 (ID: 60001044004), with total durations of 166 and 131\,ks, respectively.  We used the \textsc{nupipeline} and \textsc{nuproducts} scripts, as part of the {\it NuSTAR} Data Analysis Software (\textsc{nustardas}) package (v. 1.4.1), to extract spectral products using the calibration database (CALDB: 20150316).  Spectral products were extracted from circular source regions of 1.25\,arc min radii while background products were extracted from equal-sized circular regions separated from the source and the edges of the CCD.  After source extraction, the net exposure times for the two observations were 79 and 65\,ks, respectively.  The source is well-detected throughout the 3--79\,keV {\it NuSTAR} bandpass.  The {\it NuSTAR} spectra were binned in order to over-sample the instrumental resolution by at least a factor of 2.5 and to have a Signal-to-Noise Ratio (SNR) $> 5\sigma$ in each spectral channel.  This also ensures that our spectra contain $> 25$\,ct\,bin$^{-1}$, allowing us to use $\chi^{2}$ minimization.

{\it NuSTAR} consists of two Focal Plane Modules: FPMA and FPMB.  We analyzed data from both modules simultaneously, allowing for a floating multiplicative component to account for cross-normalization uncertainties.  Typically, the FPMB/FPMA cross-calibration normalization is $1.01 \pm 0.03$, consistent with the cross-calibration results presented in \citet{Madsen15}.  The FPMA+FPMB background-subtracted 3--79\,keV count rates of the two NuSTAR observations are $\sim$1.24 and $\sim$1.85\,ct\,s$^{-1}$, respectively. These corresponds to respective 3--79\,keV observed fluxes of $\sim$7.1 $\times 10^{-11}$ and $\sim$1.0 $\times 10^{-10}$\,erg\,cm$^{-2}$\,s$^{-1}$.

\section{Results} \label{sec:results}

Here, we begin to analyze the variability of Ark\,120.  We begin by analyzing the {\it XMM-Newton} observations.  For any subsequent spectral fits, all spectra were binned up such that there were $> 25$\,ct\,bin$^{-1}$.  Meanwhile, errors are provided at the 90\,per cent confidence level (i.e. corresponding to $\Delta \chi^{2} = 2.71$), unless otherwise stated.  A time-averaged `mean' spectrum was created by combining the 2014a, 2014b, 2014c and 2014d spectra using the \textsc{epicspeccombine}\footnote{\url{http://xmm-tools.cosmos.esa.int/external/sas/current/doc/epicspeccombine.pdf}} task, weighting the response files according to the exposure time.  All spectra were fitted with \textsc{xspec} v.12.9.0 \citep{Arnaud96}.  In our fits, neutral gas and dust is modelled by the  \textsc{tbabs} component \citep{WilmsAllenMcCray00}, where we also utilize the photoionization absorption cross-sections of \citet{Verner96}.  To account for the Galactic hydrogen column, we fixed this at a value of $1.40 \times 10^{21}$\,cm$^{-2}$ based on the measurements of \citet{Kalberla05} at the location of this source.  This value has been modified according to \citet{Willingale13} to include the impact of molecular hydrogen (H$_{\rm 2}$).  We note that the true measure of Galactic absorption and its dependency on the assumed continuum model is explored in paper I, through an analysis of the high-resolution RGS data.

\subsection{{\it XMM-Newton} variability} \label{sec:xmm_variability}

\begin{figure*}
\begin{center}
\rotatebox{0}{\includegraphics[width=17.8cm]{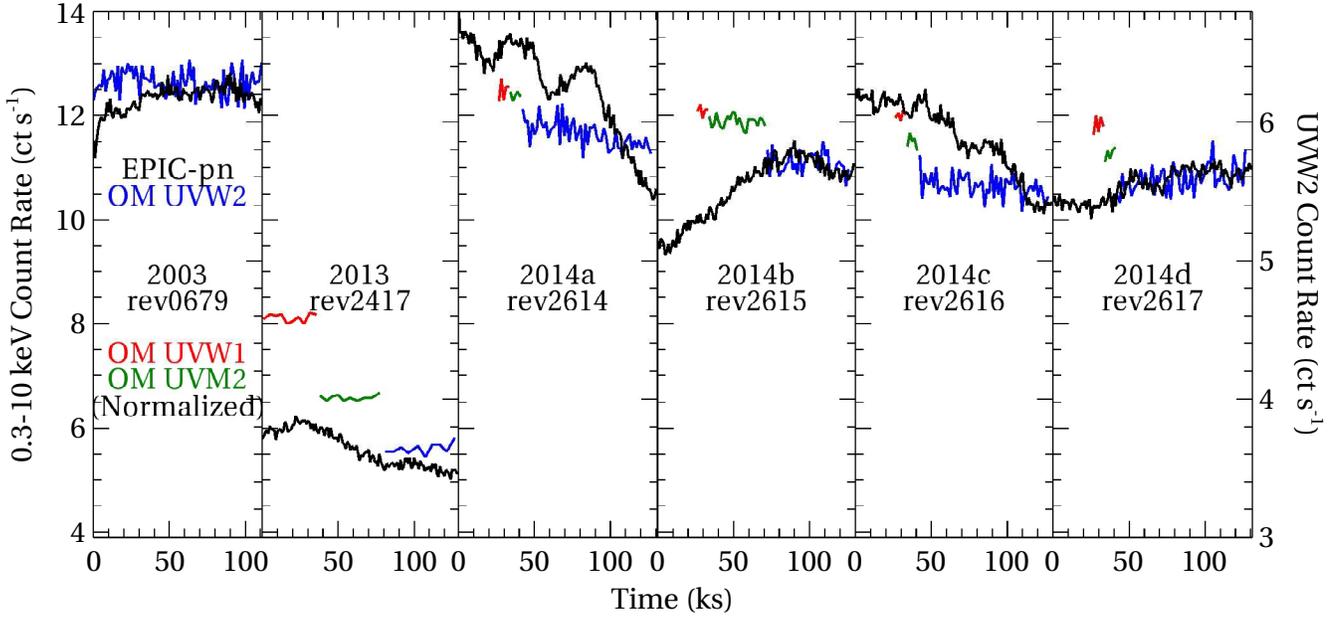}}
\end{center}
\vspace{-15pt}
\caption{The concatenated light curve of Ark\,120 obtained from all six {\it XMM-Newton} observations.  The 0.3--10\,keV EPIC-pn X-ray light curve (single pixel events only) is shown in black while monitoring data acquired with the OM's UVW2 filter are shown in blue.  Meanwhile, the UVW1 and UVM2 data are shown in red and green, respectively, and have been normalized by the UVW2 corrected count rate.}
\label{fig:pn_om_lc}
\end{figure*}

We begin by examining the broad-band {\it XMM-Newton} light curves.  These were extracted from EPIC event files using custom \textsc{idl}\footnote{\url{http://www.exelisvis.co.uk/ProductsServices/IDL.aspx}}\footnote{\url{http://www.star.le.ac.uk/sav2/idl.html}} scripts.  Background-subtraction was applied and exposure loss corrections were made, while interpolating over short telemetry drop-outs, where necessary. The 0.3--10\,keV EPIC-pn concatenated light curves for each of the six Ark\,120 observations are presented in Fig.~\ref{fig:pn_om_lc} with a time binning of 1\,ks.  The observed count rate varies by up to a factor of $\sim$2 between observations and varies by up to $\sim$30\,per cent within any individual observation (on timescales of a few tens of ks).  It is quite clear that the source was much fainter in 2013 before returning to a brighter state in 2014, more consistent with the flux level observed in 2003.

Superimposed on the EPIC-pn X-ray light curves are the UV light curves obtained from the simultaneous OM monitoring.  We note that there may exist some significant contribution to the observed nuclear light by the host galaxy --- however, this should largely remain constant.  The UVW2 data are shown in blue and clearly track the X-ray behaviour of the source on long timescales, mimicking the drop in flux in 2013 and subsequent rise in 2014.  On long timescales, the UVW2 data vary by up to $\sim$80\,per cent, although their within-observation variability is much less pronounced at just a few per cent.  In addition to the data acquired with the UVW2 filter, the UVW1 and UVM2 data are also shown in red and green, respectively.  These data have been normalized by the UVW2 count rate, such that the relative variability of the light curves remains intact.  In similar fashion to the UVW2 data, the UVW1 and UVM2 light curves are also observed to track the X-ray variations on long timescales, displaying the characteristic drop in flux in 2013.

We performed a quick test to quantify any long-term UV/X-ray correlations with the Pearson correlation coefficient, $r$.  This measures the strength of any linear relationship between two different variables.  We used the mean count rates per observation for each of the UV filters and the EPIC-pn instrument.  The correlation coefficients for the three UV filters versus the 0.3--10\,keV EPIC-pn count rates are: UVW1 $= 0.981$ ($p = 0.003$), UVM2 $= 0.977$ ($p = 0.004$) and UVW2 $= 0.971$ ($p = 0.001$).  As such, the UV/X-ray fluxes appear to be significantly correlated on long timescales (i.e. $\sim$years), a result which is significant at the $> 99$\,per cent level.  However, if we re-compute the X-ray vs UVW2 correlation coefficient using just the high-flux data (i.e. removing the datapoint from 2013), the coefficient drops to $r = 0.682$ ($p = 0.20$), indicating that the drop in flux in 2013 is driving the correlation. 

\begin{figure}
\begin{center}
\rotatebox{0}{\includegraphics[width=8.4cm]{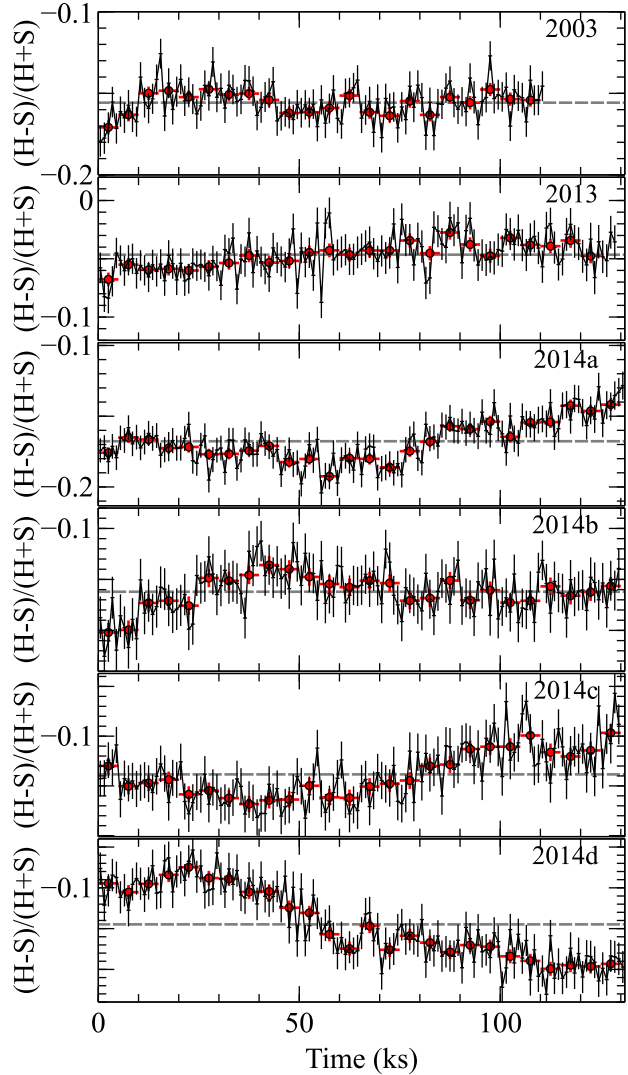}}
\end{center}
\vspace{-15pt}
\caption{The six {\it XMM-Newton} light curves of Ark\,120 plotted in terms of their hardness ratios [$(H-S)/(H+S)$], where $H$ and $S$ are the hard (1--10\,keV) and soft (0.3--1\,keV) energy bands, respectively.  The horizontal dashed lines mark the mean hardness ratio for each light curve.  Light curves are shown with 1\,ks (black) and 10\,ks (red) binning.}
\label{fig:all_lc_soft_hard_ratio}
\end{figure}

In Fig.~\ref{fig:all_lc_soft_hard_ratio}, we show the variations in the `hardness' of the six EPIC-pn light curves by plotting them according to the formula: $HR = (H-S)/(H+S)$, where $H$ and $S$ refer to the hard and soft bands, respectively (see \citealt{Park06} for a discussion of hardness ratio estimation).  Here, the soft band refers to the 0.3--1\,keV energy band while the hard band refers to the 1--10\,keV band.  It can be seen that relatively weak variability in the hardness ratio is observed within individual observations of $\lesssim$10\,per cent, generally on timescales of tens of ks.  This implies mild spectral variability within a given orbit, but this is much weaker than the between-orbit variations (either years or, in the case of the 2014 campaign, roughly $\sim$1/2\,d).

\begin{figure}
\begin{center}
\rotatebox{0}{\includegraphics[width=8.4cm]{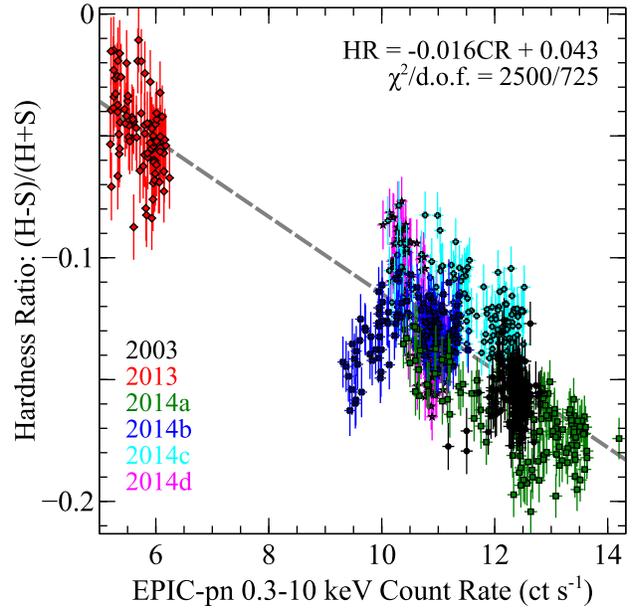}}
\end{center}
\vspace{-15pt}
\caption{Hardness ratios [($H-S)/(H+S)$] computed in 1\,ks bins from all six {\it XMM-Newton} observations of Ark\,120 and plotted against the broad-band 0.3--10\,keV count rate.  A linear trend has been fitted to the data, indicating softer-when-brighter behaviour.}
\label{fig:all_pn_1ks_hr}
\end{figure}

Additionally, in Fig.~\ref{fig:all_pn_1ks_hr}, we investigate the long-term variability of Ark\,120 by computing hardness ratios for 725 1\,ks bins obtained from all six {\it XMM-Newton} EPIC-pn observations.  As in Fig.~\ref{fig:all_lc_soft_hard_ratio}, we define the hard and soft bands to be 1--10 and 0.3--1\,keV, respectively.  We then plotted this against the broad-band 0.3--10\,keV count rate.  The long-term trend appears to be that the source displays clear softer-when-brighter behaviour.  We fitted a simple linear function of the form: $HR = aCR + b$, where $HR$ is the hardness ratio, $CR$ is the 0.3--10\,keV count rate and $a$ and $b$ are the slope and intercept, respectively.  We find best-fitting values of $a = -0.016$ and $b = 0.04$ with a fit statistic of $\chi^{2}/{\rm d.o.f.} = 2\,500/725$.  To test whether the slope is dominated by the low-flux 2013 data, we also fitted the linear trend to the 2003+2014 data only.  We again find that the best-fitting slope is negative and does, in fact, remain at the same value of $a = -0.016$, while the offset is $b = 0.05$, again indicative of softer-when-brighter behaviour of the source.  The fit statistic is $\chi^{2}/{\rm d.o.f.} = 2\,382/631$.

\begin{figure}
\begin{center}
\rotatebox{0}{\includegraphics[width=8.4cm]{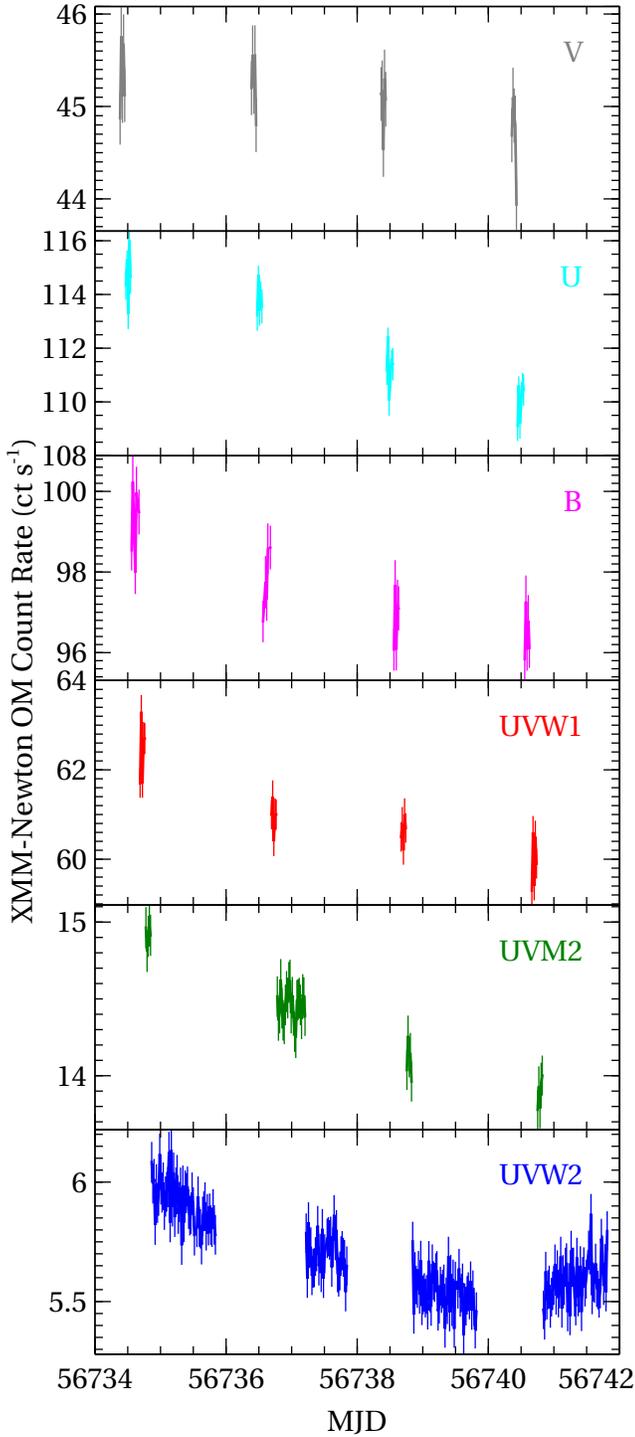}}
\end{center}
\vspace{-15pt}
\caption{The {\it XMM-Newton} OM corrected count rates obtained with V, U, B, UVW1, UVM2 and UVW2 filters, shown in the upper to lower panels, respectively.  A steady decrease in flux over this $\sim$week-long period can be observed in all filter bands.}
\label{fig:2014_om_lc}
\end{figure}

\begin{table*}
\centering
\begin{tabular}{l c c c c c c}
\toprule
{\it XMM} & \multicolumn{6}{c}{Corrected Count Rate per OM Filter (ct\,s$^{-1}$)} \\
Observation & V (5\,430\,\AA) & B (4\,500\,\AA) & U (3\,440\,\AA) & UVW1 (2\,910\,\AA) & UVM2 (2\,310\,\AA) & UVW2 (2\,120\,\AA) \\
\midrule
2003 & --- & --- & --- & --- & --- & $6.29 \pm 0.01$ \\
\midrule
2013 & --- & --- & --- & $45.97 \pm 0.05$ & $9.66 \pm 0.02$ & $3.64 \pm 0.01$ \\
\midrule
2014a & $45.31 \pm 0.13$ & $114.72 \pm 0.22$ & $99.24 \pm 0.26$ & $62.42 \pm 0.16$ & $14.91 \pm 0.06$ & $5.91 \pm 0.01$ \\
2014b & $45.27 \pm 0.14$ & $113.81 \pm 0.23$ & $97.80 \pm 0.26$ & $60.96 \pm 0.15$ & $14.46 \pm 0.02$ & $5.70 \pm 0.01$ \\
2014c & $45.04 \pm 0.14$ & $111.33 \pm 0.22$ & $96.82 \pm 0.25$ & $60.64 \pm 0.15$ & $14.11 \pm 0.06$ & $5.54 \pm 0.01$ \\
2014d & $44.69 \pm 0.13$ & $110.02 \pm 0.22$ & $96.42 \pm 0.26$ & $59.94 \pm 0.15$ & $13.89 \pm 0.06$ & $5.59 \pm 0.01$ \\
Mean (ct\,s$^{-1}$) & $45.07 \pm 0.07$ & $112.47 \pm 0.11$ & $97.57 \pm 0.13$ & $60.99 \pm 0.08$ & $14.34 \pm 0.03$ & $5.68 \pm 0.01$ \\
$F_{\rm var}$\,(per cent) & $0.6 \pm 0.1$ & $1.7 \pm 0.1$ & $1.2 \pm 0.1$ & $1.9 \pm 0.1$ & $2.0 \pm 0.1$ & $2.8 \pm 0.1$ \\
Mean Flux ($\times 10^{-15}$\,erg\,cm$^{-2}$\,s$^{-1}$\,\AA$^{-1}$) & $11.22 \pm 0.02$ & $14.51 \pm 0.01$ & $18.93 \pm 0.02$ & $29.03 \pm 0.04$ & $31.55 \pm 0.06$ & $32.46 \pm 0.03$ \\
\bottomrule
\end{tabular}
\caption{The corrected observed mean count rates for each filter utilized by the OM averaged over each observation.  The effective wavelengths of each filter are provided in parentheses.  The mean, $F_{\rm var}$ and flux values correspond to the 2014 data.  See Section~\ref{sec:xmm_variability} for details.}
\label{tab:om_count_rates}
\end{table*}

In Fig.~\ref{fig:2014_om_lc}, we show the corrected count rates obtained from each of the optical/UV filters for the 2014 data.  From upper to lower, the six panels depict the corrected count rates in the V, U, B, UVW1, UVM2 and UVW2 bands, respectively, over the $\sim$week-long period in 2014.  Each data bin corresponds to the integrated count rate obtained from an individual image.  A steady decrease in flux can be observed in each of the six filter bandpasses, with observed variations on the order of a $\sim$few per cent on the timescale of $\sim$days.  Meanwhile, there is a hint of a small increase in flux towards the end of the UVW2 light curve (observation: 2014d), suggesting a potential upturn in the shorter wavelength emission.  Indeed, the average UVW2 count rate over the first three of the 2014 {\it XMM-Newton} observations falls steadily from $\sim$5.91 to $\sim$5.70 to $\sim$5.54\,ct\,s$^{-1}$, respectively, before increasing slightly to $\sim$5.59\,ct\,s$^{-1}$ in the final observation.

\begin{figure}
\begin{center}
\rotatebox{0}{\includegraphics[width=8.4cm]{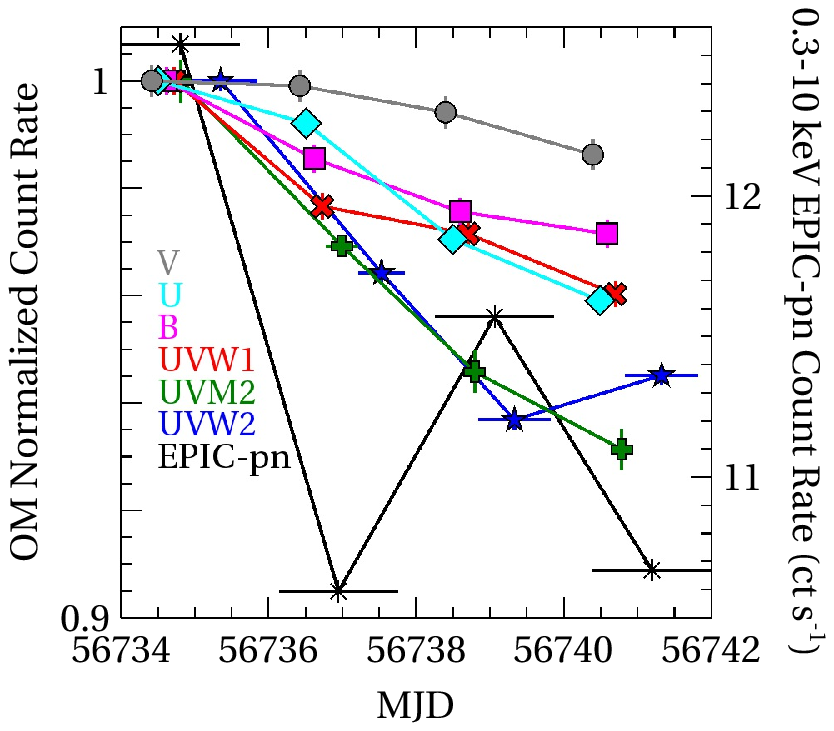}}
\end{center}
\vspace{-15pt}
\caption{The normalized {\it XMM-Newton} OM light curves, binned up by observation.  In order of decreasing wavelength, the V, U, B, UVW1, UVM2 and UVW2 bands are shown in grey, cyan, magenta, red, green and blue, respectively.  The averaged 0.3--10\,keV EPIC-pn count rates are superimposed in black.}
\label{fig:2014_om_lc_normalized}
\end{figure}

The relationship between the variability observed in the various OM bands can be better visualised in Fig.~\ref{fig:2014_om_lc_normalized}.  Here, the count rates have been averaged over each {\it XMM-Newton} observation and normalized such that the count rate in the 2014a observation is equal to 1.  This allows the relative variability between bands to be visualised.  Again, it is quite clear that the flux of the source decreases steadily in each band, with the variability in the shorter wavelength bands significantly more pronounced on timescales of $\sim$days.  For example, the UVW2 flux varies by a factor of $\sim$5--6\,per cent while the V-band flux only varies by up to $\sim$1--2\,per cent --- i.e. the UV data are relatively more variable than the corresponding optical data.

Once again, a slight upturn in flux can be observed in the UVW2 band in the final observation.  For comparison, the averaged 0.3--10\,keV EPIC-pn count rates are superimposed on the plot, illustrating that the X-ray band is significantly more variable between observations.  Intriguingly, the X-ray data also show an upturn in flux, although this occurs in 2014c, $\sim$2\,days before the UVW2 upturn.  The relationship between X-ray/UV variability is more thoroughly explored in Section~\ref{sec:swift_monitoring}, where the results of the {\it Swift} monitoring campaign are presented.

Now, the amplitude of variability across all the 2014 observations appears relatively modest at a $\sim$few\,per cent. However, this does not take into account the measurement uncertainties, which will also contribute to the total observed variance.  One method of quantifying variability is to calculate the `excess variance' (\citealt{Nandra97}, \citealt{Edelson02}, \citealt{Vaughan03}), which allows one to estimate the intrinsic source variance, and is defined as:

\begin{equation}\sigma^{2}_{\rm XS} = S^{2} - \overline{\sigma^{2}_{\rm err}}, \end{equation} 

where $S^{2}$ is the sample variance:

\begin{equation} S^{2} = \frac{1}{N-1}\sum\limits^{N}_{i=1}(x_{\rm i} - \overline{x})^{2}, \end{equation}

and $\overline{\sigma^{2}_{\rm err}}$ is the mean square error:

\begin{equation} \overline{\sigma^{2}_{\rm err}} = \frac{1}{N}\sum\limits^{N}_{i=1}\sigma^{2}_{\rm err, i}. \end{equation}

In the above equations, $x_{\rm i}$ is the observed value, $\overline{x}$ is its arithmetic mean and $\sigma_{\rm err}$ is the uncertainty on each measurement.  The excess variance then allows us to calculate $F_{\rm var}$, which is the fractional root mean square variability:

\begin{equation} \label{eq:f_var} F_{\rm var} = \sqrt{\frac{S^{2}-\overline{\sigma^{2}_{\rm err}}}{\overline{x}^{2}}}, \end{equation}

which can be expressed in terms of per cent.  As suggested by Fig.~\ref{fig:2014_om_lc_normalized}, $F_{\rm var}$ is greater for the shorter-wavelength UV emission (UVW2 $\sim 2.8$\,per cent) compared to the longer-wavelength optical emission (V $\sim 0.6$\,per cent).  These values, along with all corrected count rates (averaged per observation and across observations) are provided in Table~\ref{tab:om_count_rates}.

Finally, we obtained rough estimates of the mean observed flux in each OM bandpass by using count-rate-to-flux conversion factors provided by the \textsc{sas} team\footnote{\url{http://www.cosmos.esa.int/web/xmm-newton/sas-watchout-uvflux}}.  These are listed in Table~\ref{tab:om_count_rates} in units of $\times 10^{-15}$\,erg\,cm$^{-2}$\,s$^{-1}$\,\AA$^{-1}$.

\subsubsection{Spectral decomposition} \label{sec:spectral_decomposition}

Here, we perform a simple spectral decomposition of Ark\,120.  The broad-band X-ray spectrum is analyzed in detail in paper IV, where it is observed to be dominated by a hard X-ray power law, a significant `soft excess' at low energies and a strong Fe\,K$\alpha$ emission complex at $\sim$6--7\,keV.  A Comptonization model is favoured for the broad-band spectrum with the soft X-ray spectrum dominated by Comptonization of seed photons in a warm, optically-thick corona ($kT_{\rm e} \sim 0.5$\,keV; $\tau \sim 9$).  Meanwhile, the Fe\,K complex is well-fitted by a mildly-relativistic disc-reflection spectrum.  The finer details of the Fe\,K profile are presented in paper II, where excess components of emission both red-ward and blue-ward of the core of the line are observed to vary on different timescales, indicative of arising from short-lived hotspots on the surface, a few tens of $r_{\rm g}$ from the black hole.

\begin{figure}
\begin{center}
\rotatebox{0}{\includegraphics[width=8.4cm]{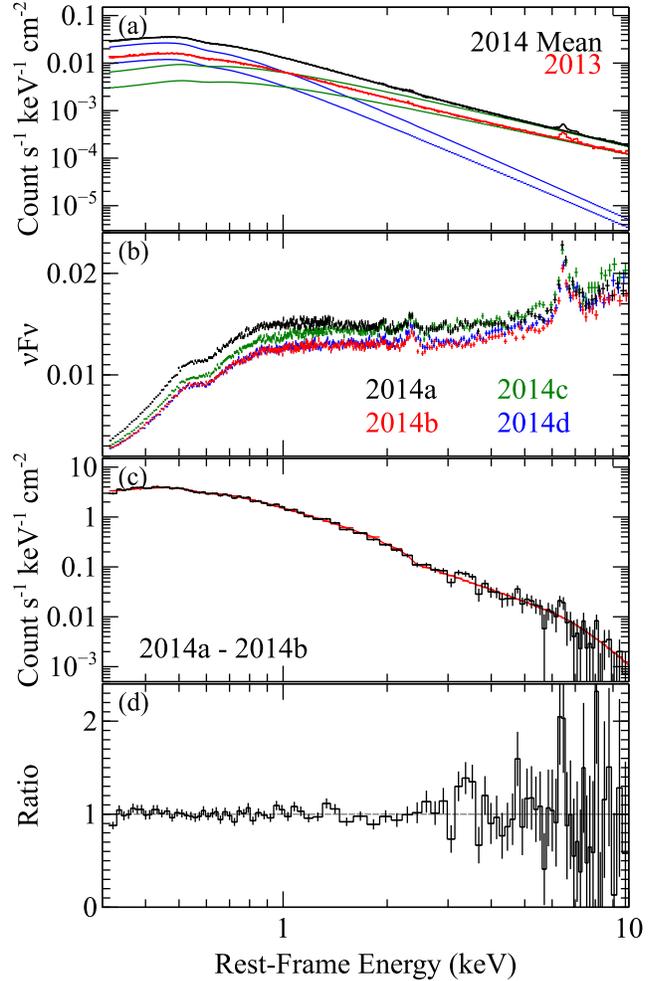}}
\end{center}
\vspace{-15pt}
\caption{Panel (a): the 2014 mean (black) and 2013 (red) EPIC-pn spectra.  The contributions of the hard and soft power laws are shown in green and blue, respectively (dashed lines = 2014 mean; dotted = 2013).  Panel (b): the ``fluxed'' EPIC-pn spectra of Ark\,120 obtained in 2014 (not corrected for Galactic absorption).  Panel (c): the `difference spectrum' obtained from subtracting the low-flux spectrum (2014b) from the high-flux spectrum (2014a).  The solid red curve represents the model fitted to the spectrum.  Panel (d): the ratio of the data-to-model residuals obtained from fitting the difference spectrum.  All spectra have been binned up for clarity.}
\label{fig:eeuf_diff}
\end{figure}

Panel (a) of Fig.~\ref{fig:eeuf_diff} shows the counts spectra of the 2014 mean and 2013 EPIC-pn data, fitted with a simple two-component baseline model consisting of a hard power law (2014: $\Gamma \sim 1.7$; 2013: $\Gamma \sim 1.5$) and a softer power law ($\Gamma \sim 3.2$) to model the soft excess, which dominates at energies $\lesssim 1.5$\,keV.  The difference between the two spectra is dominated by variations in the normalization of the two continuum components, which are both $\sim$50\,per cent lower in the 2013 epoch.  In panel (b), the broad-band spectral variations on timescales of $\sim$days are investigated as it shows all four EPIC-pn spectra obtained over the course of $\sim$1-week in 2014 ``fluxed'' against a power law with $\Gamma = 0$.  All four spectra appear to be relatively hard with some modest variations in the spectral shape between orbits\footnote{The clear residual structure at $\sim$2.2\,keV arises from inaccuracies in the calibration of the EPIC-pn energy scale, as discussed in \citet{Marinucci14}.  For detailed spectral analysis --- e.g. paper II and paper IV --- a gain shift is applied to the fits.  Details are provided in paper II.}.  

To investigate spectral variation in more detail, a series of `difference spectra' were created.  In Fig.~\ref{fig:eeuf_diff} (panel c), we show the effect of subtracting the lowest-flux spectrum (2014b) from the highest-flux spectrum (2014a) with identical binning.  The resultant spectrum is well-fitted ($\chi^{2}/{\rm d.o.f.} = 1\,763/1\,707$) with a single absorbed power law ($\Gamma = 2.45 \pm 0.10$) plus a weaker component of soft-band emission (e.g. a blackbody: $kT = 0.12 \pm 0.03$\,keV).  The steep power law dominates, suggesting that the bulk of the variability arises in the soft band.  The residuals [panel (d)] suggest that a modest component of Fe\,K emission at $\sim$6.4\,keV is varying between orbits (i.e. on timescales of $\sim$tens of ks).  

\begin{figure}
\begin{center}
\rotatebox{0}{\includegraphics[width=8.4cm]{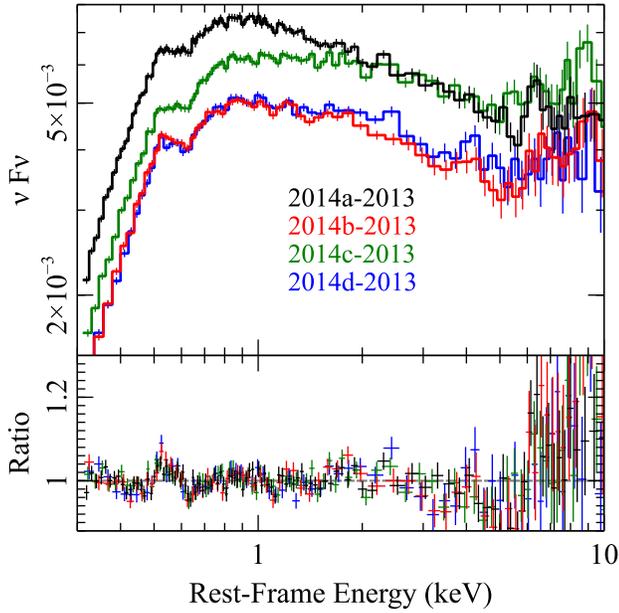}}
\end{center}
\vspace{-15pt}
\caption{Long-term EPIC-pn difference spectra having subtracted the low-flux 2013 observation from the four high-flux 2014 spectra.  The upper panel shows the ``fluxed'' difference spectra while the lower panel shows the ratio of the residuals to a fit of the continuum model of the form: \textsc{tbabs} $\times$ (\textsc{pl} $+$ \textsc{pl}).  All spectra have been binned up for clarity.  See Section~\ref{sec:spectral_decomposition} for details.}
\label{fig:2014_2013_diff}
\end{figure}

\begin{figure}
\begin{center}
\rotatebox{0}{\includegraphics[width=8.4cm]{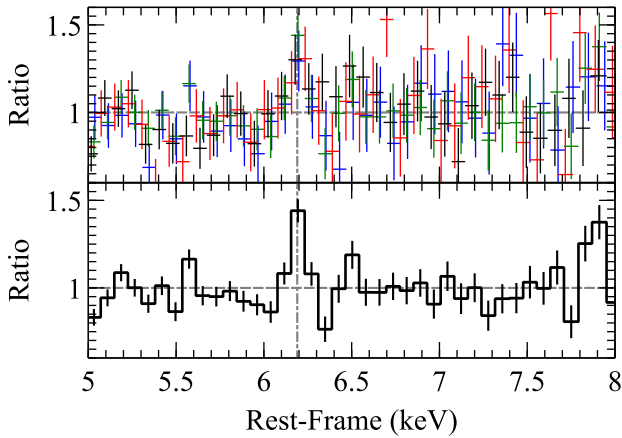}}
\end{center}
\vspace{-15pt}
\caption{The ratio of the residuals to a continuum fit of the form \textsc{tbabs} $\times$ (\textsc{pl} $+$ \textsc{pl}) to the long-term 2014/2013 EPIC-pn difference spectra from 5--8\,keV.  The upper panel shows the four individual observations: 2014a (black), 2014b (red), 2014c (green) and 2014d (blue).  Meanwhile, for extra clarity, the lower panel shows the residuals of the difference spectrum where the 2014 data have been combined into a single spectrum.  A clear excess of emission remains at $\sim$6.2\,keV.}
\label{fig:2014_2013_diff_fek}
\end{figure}

Meanwhile, Fig.~\ref{fig:2014_2013_diff} shows four longer-term difference spectra where we subtracted the low-flux 2013 EPIC-pn spectrum from each of the high-flux 2014 spectra to observe the spectral differences on the timescale of $\sim$1\,yr.  Here, the dominant power law is found to be curiously steeper than the primary power law in the time-averaged spectra, lying in the range: $\Gamma = 2.1 - 2.2$.  An additional component of soft-band emission is still required, although a second steep power law ($\Gamma =  3.5 - 4.1$) is preferable to a blackbody ($kT \sim 0.1$\,keV; $\chi^{2} / {\rm d.o.f.} = 7\,730/6\,785$).  Fig.~\ref{fig:2014_2013_diff_fek} shows the ratio of the residuals to this fit over the 5--8\,keV range, encompassing the Fe\,K band.  A clear signature of excess variable emission can be observed red-ward of the core of the Fe\,K emission complex at $\sim$6.2\,keV and is likely the variable `red wing' of the emission line observed in paper II.

\begin{figure}
\begin{center}
\rotatebox{0}{\includegraphics[width=8.4cm]{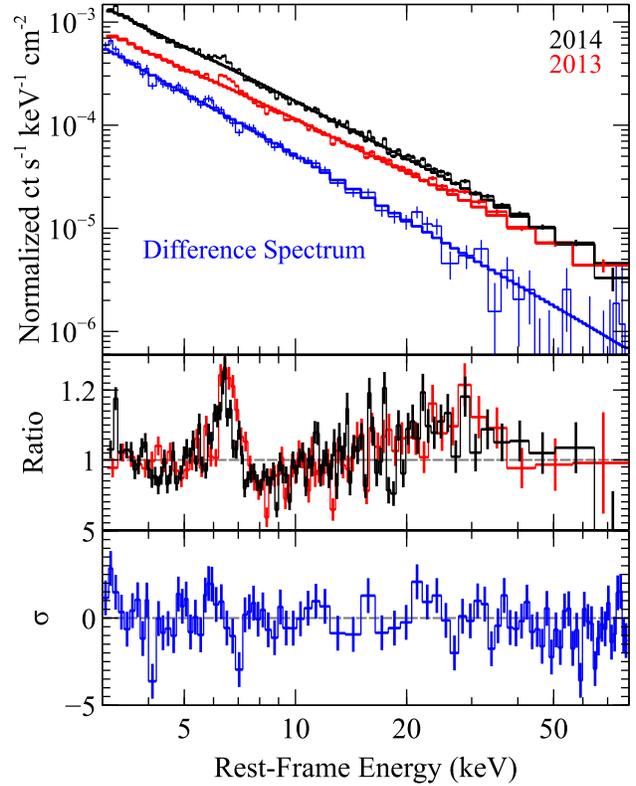}}
\end{center}
\vspace{-15pt}
\caption{Upper panel: the 2014 (mean) and 2013 broad-band {\it NuSTAR} spectra (black and red, respectively), fitted with a simple power law.  The difference spectrum between the two, also fitted with a power law, is shown in blue.  Middle panel: the ratio of the residuals to the power laws fitted to the 2014 and 2013 {\it NuSTAR} spectra.  Lower panel: the $\sigma$ residuals to the power law fitted to the difference spectrum.  The FPMA and FPMB spectra have been combined in the plot for clarity.}
\label{fig:nustar_diff}
\end{figure}

Finally, we also created difference spectra based on the 2013 and 2014 {\it NuSTAR} observations.  We fitted the FPMA and FPMB modules separately, tying all parameters together and including a variable multiplicative constant component to account for cross-normalization differences.  The {\it NuSTAR} spectra and associated difference spectrum are shown in Fig.~\ref{fig:nustar_diff}, all fitted with a simple power law.  A neutral absorber is not required by the fit as the data do not extend below 3\,keV.  This model fits the general shape well ($\chi^{2} / {\rm d.o.f.} = 934/821$), again with a reasonably steep power law: $\Gamma = 2.05 \pm 0.03$.  There is no obvious requirement for an additional continuum component at high energies suggesting that the Compton reflection is relatively stable on these timescales.  Again, a hint of excess variable emission can be observed just red-ward of the Fe\,K line.  Including a Gaussian improves the fit slightly ($\Delta \chi^{2} = 13$) with $E_{\rm c} = 6.21 \pm 0.16$\,keV.  The line is unresolved ($\sigma < 300$\,eV).  The best-fitting energy of the line again suggests that this is the signature of the variable `red wing' analysed in paper II.

\subsubsection{The fractional rms spectrum} \label{sec:rms_spectrum}

In this section we analyze the energy-dependence of the rms variability using the broad-band EPIC-pn data (e.g. \citealt{Vaughan03}).  This is known as the `rms spectrum'.  We firstly extracted a series of 25 light curves, using light-curve segments of equal length.  The energy bands were roughly spaced equally in log$(E)$ across the full 0.3--10\,keV range.  Then, for each of the energy bands and segments, the fractional excess variance was calculated ($\sigma_{\rm xs}^2$ / mean$^{2}$).  We averaged this over all segments.  Computing the square root of the fractional excess variance then provides the fractional rms.

\begin{figure}
\begin{center}
\rotatebox{0}{\includegraphics[width=8.4cm]{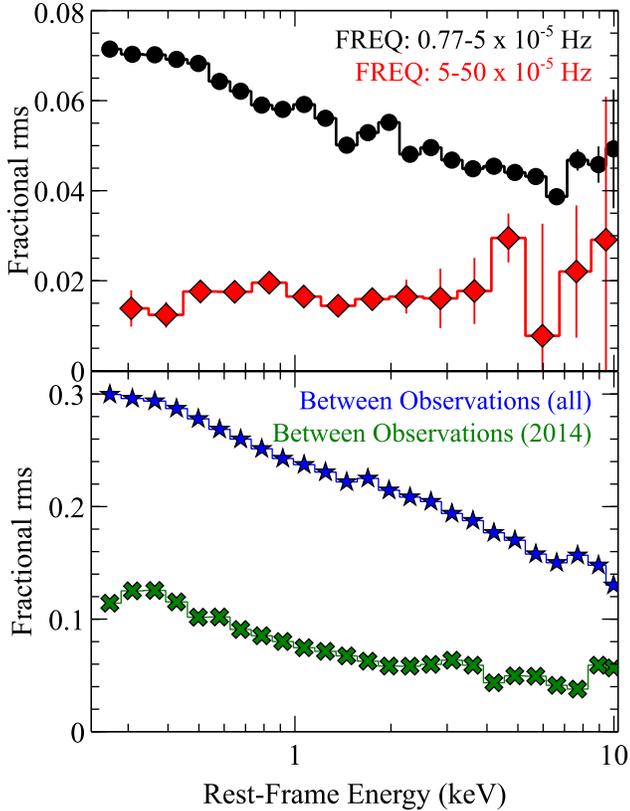}}
\end{center}
\vspace{-15pt}
\caption{The 0.3--10\,keV EPIC-pn rms spectra of Ark\,120.  Upper panel: the within-observation rms spectra, averaged over all observations acquired in 2014.  The circles (black) and diamonds (red) correspond to a lower 0.77--5 $\times 10^{-5}$ and a higher 5--50 $\times 10^{-5}$\,Hz frequency range, respectively.  Lower panel: the long-term `between-observation' rms spectra averaged over the four 2014 orbits (green crosses) and all observations (blue stars), respectively.  See Section~\ref{sec:rms_spectrum} for details.}
\label{fig:rms-spectrum}
\end{figure}

To investigate the rms behaviour on within-observation timescales, we computed the spectrum using two different frequency bands: $\sim$0.77--5 $\times 10^{-5}$\,Hz ($\Delta t = 10$\,ks; 130\,ks segments)\footnote{Here, the upper frequency bound is set by the Nyquist frequency: $\nu_{\rm N} = 1/2t$.} and 5--50 $\times 10^{-5}$\,Hz ($\Delta t = 1$\,ks; 20\,ks segments), and averaged over all four orbits from 2014 (see Fig.~\ref{fig:rms-spectrum}: upper panel).  Given that the high-frequency rms spectrum becomes noisy at high energies, we have binned up to 15 bins to increase the signal.  At high frequencies, the rms variability is low and is approximately flat across the full energy range.  Meanwhile, at lower frequencies, not only is the variability stronger, but there is a gentle rise in fractional variability towards lower energies, suggesting that the soft band dominates the observed variability.  In both spectra there is a small hint of a modest drop in fractional rms in the $\sim$6--7\,keV band, which may arise from a component of quasi-constant Fe\,K emission which may not be responding to continuum variations on these timescales.       

Finally, we computed the fractional rms spectrum on long-term `between-observation' timescales.  This is an alternative approach to viewing the spectral variations observed in Fig~\ref{fig:eeuf_diff}.  As shown in the lower panel of Fig.~\ref{fig:rms-spectrum}, the variability profile is relatively smooth and significantly stronger at low energies on these observed timescales of $\gtrsim$\,days.  The relatively steep shape of the spectrum is reminiscent of the steep difference spectrum shown in Fig.~\ref{fig:eeuf_diff} (middle panel; $\Gamma \sim 2.5$).  The blue curve in Fig.~\ref{fig:rms-spectrum} shows the effect of including the EPIC-pn data from the 2003 and 2013 epochs.  Now, the fractional rms variability is much stronger with a steeper rise towards low energies, predominantly caused by the low-flux epoch in 2013.

\subsubsection{The covariance spectrum} \label{sec:covariance}

Here, we further investigate the energy-dependence of the spectral variations through the `covariance spectrum'.  This is cross-spectrally analogous to the rms spectrum and, with careful choice of a reference band, can be used to reveal the spectral shape of correlated components in a given frequency range, $\Delta \nu$.  We follow the method first outlined in \citet{WilkinsonUttley09}, where the covariance spectrum was applied to the black hole X-ray binary system GX\,339-4.  The covariance spectrum has been used in a series of subsequent analyses of variable XRBs and AGN --- e.g. \citet{Uttley11}, \citet{MiddletonUttleyDone11}, \citet{CassatellaUttleyMaccarone12}, \citet{Kara13}.  Also see \citet{Uttley14} for a review.

We calculate the covariance spectrum using two separate reference bands where the absolute variability is high: a continuum-dominated band from 1--4\,keV and a soft band, dominated by the soft excess, from 0.3--1\,keV.  As per the fractional rms spectra in Section~\ref{sec:rms_spectrum}, we investigate the covariance spectra over two separate frequency bands: $\sim$0.77--5 $\times 10^{-5}$\,Hz and 5--50 $\times 10^{-5}$\,Hz, averaged over all four orbits from 2014.  Again, we bin up the higher-frequency spectra to 15 equally-logarithmically-spaced bins to increase the signal at high energies.  Note that to avoid contamination from the Poisson errors arising from energy bins duplicated in the reference band, we always remove the band of interest in the cases where it overlaps with the reference band.  Finally, we produce covariance spectra in units of count rate by plotting the normalized covariance, given by: $\sigma^{2}_{\rm cov, norm} = \sigma^{2}_{\rm cov} / \sqrt{\sigma^{2}_{\rm xs,y}}$, where $y$ refers to the reference band\footnote{Here, $\sigma^{2}_{\rm cov} = \frac{1}{N-1}\Sigma^{N}_{i=1}(X_{i}-\overline{X})(Y_{i}-\overline{Y})$.}.  The errors are calculated on the normalized covariance as in \citet{WilkinsonUttley09}.

\begin{figure*}
\begin{center}
\rotatebox{0}{\includegraphics[width=17.4cm]{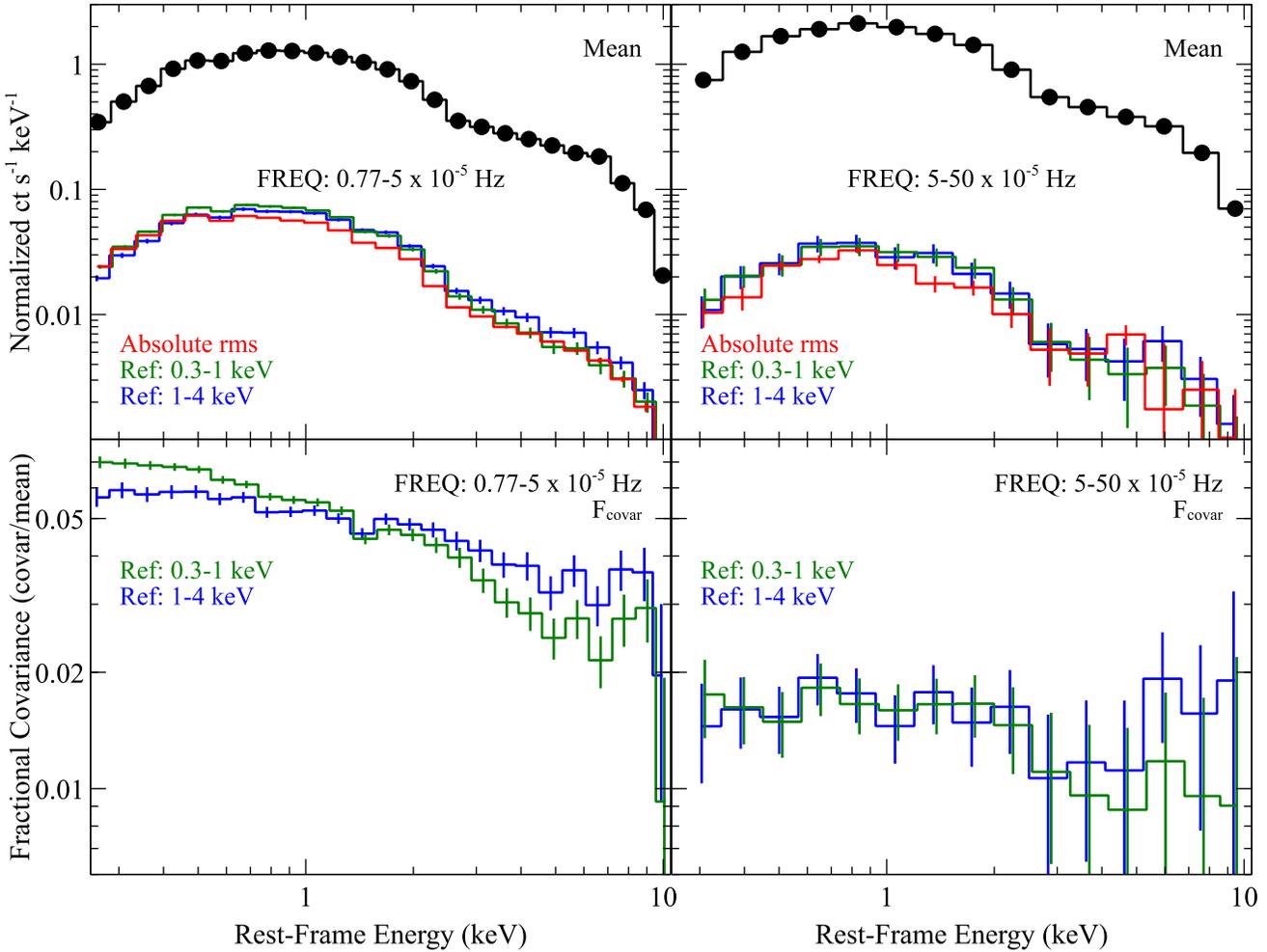}}
\end{center}
\vspace{-15pt}
\caption{Upper panels: the 0.3--10\,keV EPIC-pn covariance spectra of Ark\,120 computed against two reference bands: 0.3--1\,keV (green) and 1--4\,keV (blue).  The left-hand panels show the lower-frequency spectra (0.77--5 $\times 10^{-5}$\,Hz) while the right-hand panels show the higher-frequency spectra (5--50 $\times 10^{-5}$\,Hz).  The mean and absolute rms spectra are also shown in black and red, respectively, all with identical binning.  Lower panels: the fractional covariance spectra (i.e. covariance/mean).  Note that a slight offset on the x-axis has been applied to the covariance spectra to avoid overlapping error bars and improve clarity.  See Section~\ref{sec:covariance} for details.}
\label{fig:covariance}
\end{figure*}

The covariance spectra of Ark\,120 are shown in Fig.~\ref{fig:covariance}.  In addition, we show the `absolute rms' variability spectra (upper panel; red), which the covariance spectra should tend to as the coherence, $\gamma$, of the two light curves approaches 1 (i.e. when the variability of one band perfectly predicts the variability of the other --- see Section~\ref{sec:x-ray_time_lags} for more details).  In both frequency ranges, above $\sim$2\,keV, the covariance spectra follow a roughly power-law shape, which is mildly softer than the mean spectrum at low frequencies ($\Gamma \sim 1.9$) but becomes somewhat harder at higher frequencies, recovering the mean time-averaged value (i.e. $\Gamma \sim 1.7$), when fitting from 2--10\,keV.  The fact that the long-timescale covariance spectrum appears softer than the short-timescale covariance spectrum suggests that lower energies display relatively stronger variability on long timescales.  This can be observed in the lower panels of Fig.~\ref{fig:covariance}, where we plot the fractional covariance spectra --- i.e. by dividing by the mean spectrum, which is shown in the upper panels.  This fractional covariance also allows us to estimate the contribution to the overall variability from each component, revealing that the low-frequency variations are more dominant.  At lower energies, where the soft excess dominates, the covariance is a factor of a few stronger than it is at energies where the power law dominates, again suggesting that the soft excess in Ark\,120 is more variable than the power-law emission.

\subsubsection{The power spectrum} \label{sec:psd}

\begin{figure}
\begin{center}
\rotatebox{0}{\includegraphics[width=8.4cm]{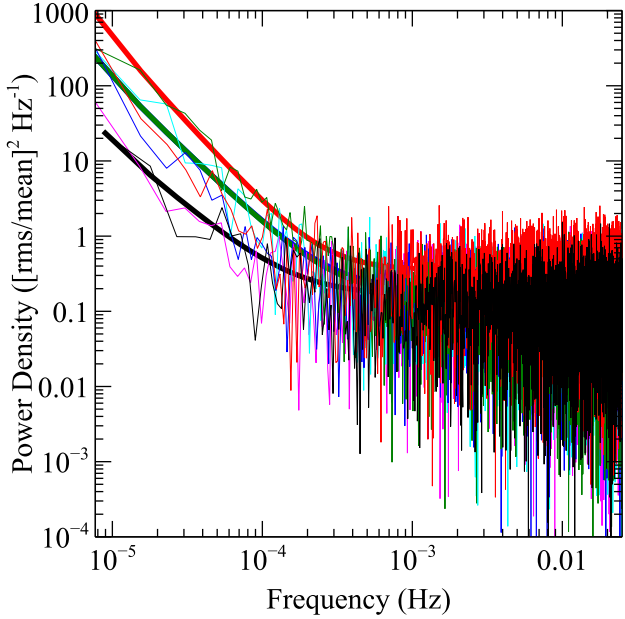}}
\end{center}
\vspace{-15pt}
\caption{The 0.3--10\,keV PSD of Ark\,120.  These are ``raw'' unbinned periodograms estimated from each of the 6 EPIC pn observations from 2003 (black), 2013 (red) and 2014 (green, blue, cyan and magenta).  The thick curves represent the fitted power-law $+$ constant model (equation~\ref{eq:psd_pl}) described in Section~\ref{sec:psd}.}
\label{fig:psd}
\end{figure}

In this section we investigate the PSD.  This estimates the temporal frequency-dependence of the power of variability.  For each of the six {\it XMM-Newton} observations of Ark\,120, we computed background-subtracted EPIC-pn light curves.  The time resolution was $\Delta t = 20$\,s.  Then, we computed periodograms from each observation in (rms/mean)$^{2}$ units (\citealt{Priestly81}; \citealt{PercivalWalden93}; \citealt{Vaughan03}; also see Fig.~\ref{fig:psd}).  As the observations are long (i.e. $\sim$130\,ks), we can estimate the PSD down to low frequencies; i.e. $\sim$7.7 $\times 10^{-6}$\,Hz.

We used \textsc{xspec} to fit the ``raw'' periodograms, initially with a model consisting of a power law plus a constant (see \citealt{Gonzalez-MartinVaughan12} for further details):

\begin{equation} \label{eq:psd_pl} m(\nu) = N\nu^{-\alpha} + C, \end{equation}

where $\alpha$ is the spectral index, $N$ is the normalization of the power law, and $C$ is an additive constant with a slope of zero in order to model high-frequency Poisson noise.  Typically, the Poisson noise is observed to dominate at frequencies $\gtrsim$3--4$\times 10^{-4}$\,Hz.  We fitted the three epochs (2003, 2013, 2014) separately, although we fitted all four contiguous observations from 2014 simultaneously.  In the latter case, we tied each parameter together between the observations apart from the normalization of the additive constant component in order to allow for variations in the Poisson noise level arising from variable count rates in the different orbits.  We found the best-fitting model parameters by minimizing the Whittle statistic, $S$:

\begin{equation} \label{eq:whittle} S = 2 \sum^{N/2}_{i=1} \left\{ \frac{y_{\rm i}}{m_{\rm i}} + {\rm ln}m_{\rm i} \right\}, \end{equation}

where $y_{\rm i}$ is the observed periodogram and $m_{\rm i}$ is the model spectral density at Fourier frequency, $\nu$ (see \citealt{Vaughan10, BarretVaughan12}).  We estimated 90\,per cent confidence intervals on each parameter by determining the set of parameter values for which $\Delta S = S(\theta) - S_{\rm min} \leq 2.7$ (where $\Delta S$ can be approximated to $\Delta \chi^{2}$).

Upon fitting with equation~\ref{eq:psd_pl}, the 2003, 2013 and 2014 epochs have 2\,785, 3\,233 and 13\,086 degrees of freedom, respectively.  Fitting the three epochs from 0.3--10\,keV results in similar best-fitting parameters with $\alpha = 1.65^{+0.42}_{-0.32}$, $2.20^{+0.55}_{-0.42}$ and $1.90^{+0.15}_{-0.13}$, respectively.  These values along with the power-law normalizations and Whittle statistics are shown in Table~\ref{tab:psd}.  We note that some values (e.g. $N$ for the 2013 epoch) are unconstrained due to the low number of bins above the Poisson noise level.  These are marked with asterisks in Table~\ref{tab:psd}.  The fitted PSDs are shown in Fig.~\ref{fig:psd}.  Meanwhile, a long-term `time-averaged' fit by including all six observations and tying $\alpha$ between all observations but allowing $N$ and $C$ to vary between datasets resulted in a best-fitting slope of $\alpha = 1.94^{+0.13}_{-0.12}$ and a Whittle statistic of $-472\,560$ for $19\,083$ degrees of freedom.  Truncating the PSDs at $10^{-3}$\,Hz to allow the low-frequency `red-noise slope' to drive the fit results in fits which are consistent with those reported in Table~\ref{tab:psd}.

\begin{table}
\centering
\begin{tabular}{l c c c c}
\toprule
\multirow{2}{*}{Epoch} & Power law & \multicolumn{3}{c}{Energy Range (keV)} \\
& (Statistic) & 0.3--10 & 0.3--2 & 2--10 \\
\midrule
& $\alpha$ & $1.65^{+0.42}_{-0.32}$ & $1.71^{+0.54}_{-0.37}$ & $1.90^{+0.68}_{-0.50}$ \\
2003 & log($N$) & $-7.1^{+2.1}_{-1.7}$ & $-7.5^{+0.6}_{-2.3}$ & $-7.8^{+0.5}_{-3.2}$ \\
& Whittle ($S$) & $-69\,318$ & $-67\,905$ & $-61\,254$ \\ 
\midrule
& $\alpha$ & $2.20^{+0.55}_{-0.42}$ & $2.02^{+0.48}_{-0.35}$ & $2.09^{+0.82}_{-0.82}$ \\
2013 & log($N$) & $-9.0^{*}$ & $-8.0^{+0.8}_{-2.1}$ & $-8.2^{*}$ \\
& Whittle ($S$) & $-76\,322$ & $-74\,272$ & $-67\,896$ \\
\midrule
& $\alpha$ & $1.90^{+0.15}_{-0.13}$ & $1.96^{+0.16}_{-0.14}$ & $2.09^{+0.32}_{-0.26}$ \\
2014 & log($N$) & $-7.5^{+0.5}_{-0.5}$ & $-7.6^{+0.5}_{-0.7}$ & $-8.4^{+0.9}_{-1.4}$ \\
& Whittle ($S$) & $-327\,042$ & $-319\,701$ & $-290\,685$ \\
\midrule
\multirow{2}{*}{All} & $\alpha$ & $1.94^{+0.13}_{-0.12}$ & $2.01^{+0.15}_{-0.14}$ & $2.10^{+0.28}_{-0.24}$ \\
& Whittle & $-472\,560$ & $-461\,946$ & $-419\,827$ \\
\bottomrule
\end{tabular}
\caption{The best-fitting parameters of the model fitted to the Ark\,120 PSD (i.e. a simple power law) in the 0.3--10, 0.3--2 and 2--10\,keV energy ranges using the {\it XMM-Newton} EPIC-pn.  Values marked with asterisks denote that we were unable to constrain the 90\,per cent confidence intervals.  We quote the Whittle statistic, $S$, for each model.  See Section~\ref{sec:psd} for details.}
\label{tab:psd}
\end{table}

We also fitted equation~\ref{eq:psd_pl} to PSDs which were generated in `soft'  and `hard' energy bands (0.3--2\,keV and 2--10\,keV, respectively) --- however, we found no significant dependence of the PSD on energy with the best-fitting values remaining consistent within the measurement uncertainties.  The best-fitting values are presented in Table~\ref{tab:psd}.  Likewise, fitting the four orbits from 2014 independently reveals that there are no significant differences between orbits.

In terms of Fourier-based PSD analysis, there are two primary forms of bias which may affect the results: `aliasing' and `leakage' (see \citealt{UttleyMcHardyPapadakis02, VaughanFabianNandra03, Gonzalez-MartinVaughan12} and reference therein).  The former occurs when observations contain gaps and high-frequency variability power above the Nyquist frequency ($\nu_{\rm Nyq} = 1/2\nu_{\rm j}$, where $\nu_{\rm j}$ is the highest observed frequency) folds back in, thus distorting the observed PSD.  As the present data are contiguously sampled, the effect of aliasing is negligible.  Meanwhile, leakage occurs when there is significant variability power at frequencies below the lowest observed frequency and can have the effect of imposing additional long-term trends on shorter observed lightcurves.  Leakage may be important for intrinsically steep PSD slopes and may significantly distort the periodogram if it is steep (e.g. $\alpha \sim 2$) at or below the lowest observed frequency (in this instance, $\nu \sim 7.7 \times 10^{-6}$\,Hz).  This can have the effect of reducing the sensitivity to features such as high-frequency bends and quasi-periodic oscillations while biasing intrinsically steep slopes to $\alpha \approx 2$ (see \citealt{DeeterBoynton82, UttleyMcHardyPapadakis02, VaughanFabianNandra03, Gonzalez-MartinVaughan12} for further details).
The simplest method to reasonably recover a better representation of the true PSD spectral index is discussed in \citet{Fougere85} and is known as `end-matching'\footnote{Red noise leakage is manifested as sinusoidal trends with periods longer than the duration of the light curve.  End-matching subtracts a linear trend (i.e. one part of the longest-duration sinusoids) and, as such, will remove some fraction, but not all, of the red-noise leaked power.}. This involves computing a linear trend ($y = mx + c$, where $y$ is the observed flux / count rate and $x$ is time) that joins the first ($y_{1}$) and last ($y_{N}$) datapoint in each light curve. This linear function is then subtracted from the observed light curve (i.e. such that $y_{1} = y_{N}$) before the mean is returned to its original value. We have end-matched the 0.3--10\,keV EPIC-pn light curves for Ark\,120 for each of the 2003, 2013 and 2014 epochs and re-fitted equation~\ref{eq:psd_pl} to the periodograms, finding $\alpha = 1.77^{+0.36}_{-0.28}$, $1.74^{+0.30}_{-0.24}$ and $2.20^{+0.25}_{-0.21}$, respectively. The 2003 and 2014 epochs now have
somewhat steeper slope estimates (although still largely consistent within the 90\,per cent uncertainties), while the 2013 epoch slope is estimated to be slightly flatter after end-matching.  However, we note that the light curve from 2013 displays only a roughly linear decreasing trend --- as such, little variability remains after end-matching.  As before, we find no significant dependence of the slope on energy and no significant 2014 inter-orbit variability.  Finally, we still find no evidence for a bend in the PSD after end-matching.

We now extend the \textit{XMM-Newton} PSDs to lower temporal frequencies by constructing broad-band, multi-segment PSDs and including long-term (days--years) monitoring data from \textit{Swift}-XRT (see below) and/or \textit{Rossi X-ray Timing Explorer} (\textit{RXTE})-PCA (Proportional Counter Array) campaigns.  We measured PSDs in three bands: hard (2--10\,keV; \textit{XMM-Newton} + \textit{Swift} + \textit{RXTE}), soft (0.3--2\,keV; \textit{XMM-Newton} + \textit{Swift}), and total (0.3--10\,keV; \textit{XMM-Newton} + \textit{Swift}).  To account for biasing/sampling effects (aliasing and red noise leakage) and to obtain proper errors on PSD points, we employ the classic ``PSRESP'' method used by \citet{UttleyMcHardyPapadakis02}, \citet{Markowitz03}, etc.  The reader is referred to those publications for details.

For each PSD, we include all six {\textit XMM-Newton} EPIC light curves, re-binned to $dt = 500$\,s, and the \textit{Swift}-XRT 6-month light curves as described above.  Ark\,120 was monitored with the \textit{RXTE} PCA once every three days from 1998 Feb 24 -- 2000 Apr 28 (MJD: 50\,868--51\,662) - a duration of 794 days.  However, there were two large gaps due to Sun-angle constraints: MJD 50\,931--51\,026 and 51\,278--51\,425, comprising 30.5\,per cent of the total duration.   We thus used two PCA light curves lasting MJD 51\,026--51\,278 (252\,d) and 51\,425--51\,662 (237\,d).  We extracted count-rate light curves following standard PCA extraction procedures (e.g. see \citealt{RiversMarkowitzRothschild13}) and constructed PSDs using Discrete Fourier Transforms (DFTs).  Following \citet{PapadakisLawrence93} and \citet{Vaughan05}, we  logarithmically binned the periodogram by a factor of $\sim$1.4 in $f$ ($\sim$0.15 in the logarithm), with the three lowest temporal frequency bins widened to each accommodate three periodogram points in order to produce the observed PSD, $P$($f$).  The resulting PSDs span the temporal frequency ranges: $1.0 \times 10^{-7} - 2.4 \times 10^{-6}$ and $1.3 \times 10^{-5} - 8.1 \times 10^{-4}$\,Hz (soft and total bands, respectively) with the hard band extending down to $7.9 \times 10^{-8}$\,Hz, but yielding a gap at $2.4 \times 10^{-6} - 1.3 \times 10^{-5}$\,Hz.

We tested two very simple models only: an unbroken power-law model (equation~\ref{eq:psd_pl}) and a sharply broken power-law:

\begin{equation} m(\nu) =  
\begin{cases}N\nu^{-\beta} + C & \textrm{for } \nu > \nu_{\rm br} \\
N\nu^{-\gamma} + C & \textrm{for } \nu \leq \nu_{\rm br}. \\
\end{cases}
\label{eq:broken_pl}
 \end{equation}

Here, $N$ represents normalization factors and $C$ denotes the constant level of power at high temporal frequencies due to Poisson noise, obtained by either eqn.\ A2 of \citet{Vaughan03} or, in the case of EPIC data, empirically from unbinned periodograms at high temporal frequencies.  Meanwhile, the frequency at which the power law slope breaks is denoted by $\nu_{\rm br}$.  More complex models such as a slow bend cannot be well constrained by the data.  When calculating $\chi^2{\rm dist}$, we ``deweighted'' the \textit{XMM-Newton} PSD points by dividing their contributions by 7 as to avoid biasing the fit towards the highest temporal frequencies.  Errors on PSD model parameters correspond to 1$\sigma$ for one interesting parameter, and correspond to values 1$\sigma$ above the rejection probability, $R$, for the best-fit model on a Gaussian probability distribution, following \citet{Markowitz03}.

\begin{figure}
\begin{center}
\rotatebox{-90}{\includegraphics[width=12.4cm]{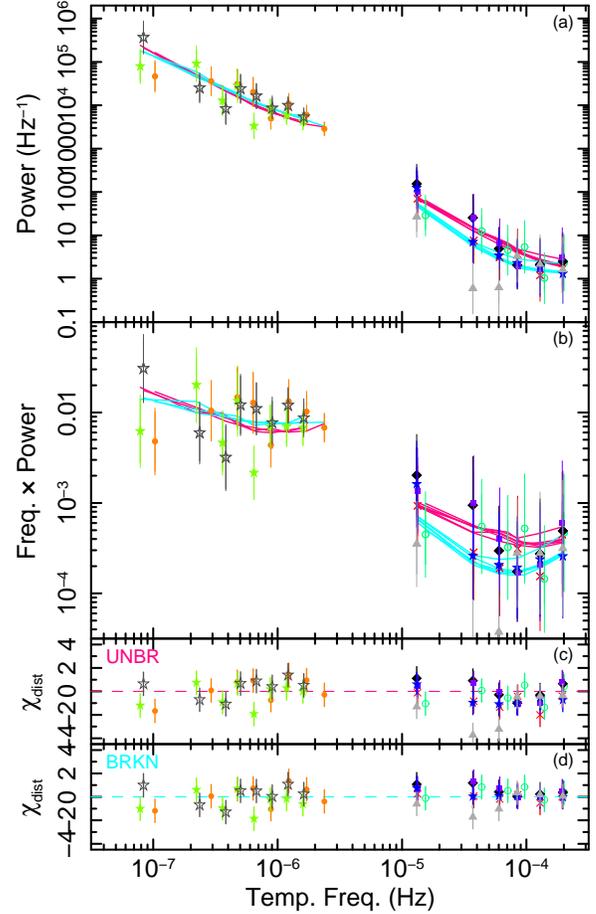}}
\end{center}
\vspace{-5pt}
\caption{The PSD of Ark\,120 in $P$($f$) space (panel a) and $f$ $\times$ $P$($f$) space (panel
b) to visually accentuate the flattening of the power-law slope towards $\sim$-1 in $P$($f$) space [$\sim$0 in $f$ $\times$ $P$($f$) space] at lower temporal frequencies.  At low temporal frequencies,
green and grey denote the two {\it RXTE} segments while orange points denote the {\it Swift} segment.  The seven {\it XMM-Newton} segments lie at higher temporal frequencies.  Red and cyan solid lines represent the best fit unbroken and broken power-law models, respectively, folded through the ``response'' of the sampling window and with Poisson noise added.  Panels c and d show the $\chi^2_{\rm dist}$ residuals for the unbroken and broken power-law models,
respectively.}
\label{fig:psd_alex}
\end{figure}

The results for the unbroken and broken power-law models for each band are listed in Table~\ref{tab:psd_results_alex}, while the PSD is plotted in Fig.~\ref{fig:psd_alex}.  $L_{\rm UNBR}$ and $L_{\rm BRKN}$ are the likelihoods of model acceptance, each calculated as $1 - R$, where $R$ is the rejection probability.  The best-fitting unbroken-power-law models yield PSD slopes, $\alpha$, with values 1.60--2.00, though formally the slopes are consistent with lower limits in the range: 1.40--1.55.  The best-fitting broken-power-law models are superior fits compared to the unbroken power law, and yield extremely small rejection probabilities: $R < 5$\,per cent.  The best-fitting break frequencies have values $\nu_{\rm br}= 2.0 - 4.0 \times 10^{-6}$\,Hz, but errors on $\nu_{\rm br}$ are large, typically $\pm \sim 0.7$ in the logarithm, due to the break's falling in the gap in temporal frequency coverage.  Best-fitting power-law slopes above the break, $\beta$, span 2.6--3.4, but are formally consistent with lower limits of 1.6--1.9.  Below the break, best-fitting values of $\gamma$ are near 1.0, with an average error of $\pm0.8$.

\begin{table*}
\centering
\begin{tabular}{l c cc c c c c c c c}
\toprule
&  \multicolumn{2}{c}{Unbroken P.L.}  &  \multicolumn{4}{c}{Broken P.L.                        } &                           \\
&  \multicolumn{2}{c}{model (UNBR)}  & \multicolumn{4}{c}{model (BRKN)} &   \\
Band (keV)           & $\alpha$ &  Rej.\ Pr.  &  $\gamma$     & log($\nu_{\rm br}$, Hz)  & $\beta$  & Rej.\ Pr. & $L_{\rm BRKN}/L_{\rm UNBR}$   \\
\midrule
Hard (2--10)  & $>1.40$  & 0.715 & $1.3^{+0.6}_{-0.6}$        & --($5.7^{+0.4}_{-1.1}$) & $>1.6$ & 0.048 & $0.952/0.295=3.2$\,$\sigma$ \\
Soft (0.3--2) & $>1.45$  & 0.786 & $0.9^{+0.8}_{-0.6}$ & --($5.6^{+0.4}_{-0.6}$) & $>1.9$ & 0.048 & $0.952/0.214=4.5$\,$\sigma$ \\
Total (0.3--10) & $>1.55$  & 0.558 & $1.2^{+0.9}_{-1.0}$  & --($5.4^{+0.8}_{-1.1}$) & $>1.8$ & 0.002 & $0.998/0.442=2.3$\,$\sigma$ \\
\bottomrule
\end{tabular}
\caption{Best-fitting model parameters obtained from the full {\it XMM-Newton} $+$ {\it Swift} $+$ {\it RXTE} PSDs.  The unbroken-power-law (UNBR) model has a power-law slope, $\alpha$, while the broken-power-law (BRKN) model has a low-frequency slope, $\gamma$, a high-frequency slope, $\beta$, and a PSD break frequency, $\nu_{\rm br}$.  The rejection probability (`Rej. Pr.') gives the ratio of the likelihoods of acceptance, which denotes the significance in the improvement of the BRKN fit.}
\label{tab:psd_results_alex}
\end{table*}

For completeness, we also measured the hard band PSD using the full-duration \textit{RXTE} PCA light curve with gaps (``full'' method).  The advantage is that the binned PSD extends down to $10^{-7.6}$\,Hz, a factor of 0.5 in the logarithm lower than using two light curves (``indiv.'' method).  However, the ``full'' method yielded only one extra binned PSD point below $10^{-7.1}$\,Hz (the lowest frequencies
probed by the ``indiv.'' method), and provided no significant additional constraint on the broadband PSD model shape.  Furthermore, the linear interpolation across gaps when measuring the DFT can introduce a dearth of high-temporal frequency variability for large gaps and slightly steepen the PSD slopes for both observed and simulated light curves.  In this case, the results obtained using the ``full'' method were completely consistent with the ``indiv.'' method: the best-fitting unbroken-power-law model had $\alpha > 1.30$; the best-fitting broken-power-law model had best-fit parameters: log($\nu_{\rm br}$\,Hz) = $-$($5.7^{+0.3}_{-1.0}$), $\beta > 1.6$, and $\gamma=1.2\pm0.5$, with $L_{\rm BRKN}/L_{\rm UNBR}$ = $0.976/0.356=2.7\sigma$.

\subsubsection{X-ray time lags} \label{sec:x-ray_time_lags}

Given that Ark\,120 is a source which displays relatively strong X-ray variability, we investigated the nature of frequency-dependent X-ray time delays between energy bands by employing Fourier methods.  By following the methods described in \citet{VaughanNowak97}, \citet{Nowak99}, \citet{Vaughan03} and \citet{Uttley11}, we calculated the cross-spectral products in two distinct broad energy bands by splitting each of the light curves into a number of identical-length segments and computing the DFT for each.  We then combined these to form auto- and cross-periodograms, averaging over all segments, allowing us to estimate the coherence, phase and time lags between the two energy bands as a function of frequency.

We initially focus on the EPIC-pn data acquired in 2014.  Given the $\sim$130\,ks length of each of the four observations, we used 65\,ks segments, providing us with 8 measurements at the lowest frequencies and allowing us to access frequencies down to $\nu \sim 1.5 \times 10^{-5}$\,Hz.  A relatively common choice of energy bands for this type of analysis is to use a `soft band' from 0.3--1\,keV and a harder, continuum-dominated band from 1--4\,keV.  As such, we opt to use these two broad energy bands here.  We used light curves extracted with time bins of size $\Delta t = 100$\,s and averaged over contiguous frequency bins, which spanned a factor of $1.1$ in frequency.

\begin{figure}
\begin{center}
\rotatebox{0}{\includegraphics[width=8.4cm]{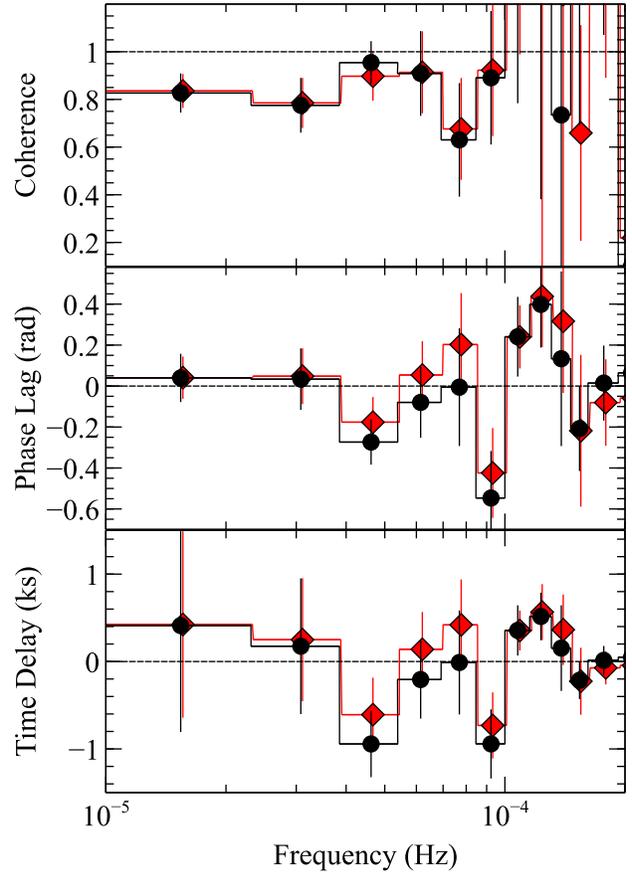}}
\end{center}
\vspace{-15pt}
\caption{Cross-spectral products computed between the 0.3--1 and 1--4\,keV energy bands using 65\,ks segments and logarithmic frequency binning.  The 2014 EPIC-pn data are shown with in black (circles) while the red (diamonds) curves show the results when including the data from 2003 and 2013.  Upper panel: the coherence between the two energy bands.  Middle panel: the phase lag.  Lower panel: the time lag, where a positive lag denotes the harder band lagging behind the softer band.  A slight offset has been applied to the combined data (red diamonds) on the x-axis to avoid overlapping error bars.}
\label{fig:lags}
\end{figure}

The cross-spectral products are shown in Fig.~\ref{fig:lags}.  The upper panel shows the coherence, $\gamma$, which is an important factor in assessing the reality of any measured time lags.  This is calculated from the magnitude of the cross-periodogram, as described in \citet{VaughanNowak97}.  It it a measure of the degree of linear correlation between the two energy bands, whose value should lie in the range $0 - 1$.  A coherence of $0$ would imply no correlation while a coherence of $1$ would mean that the variation in one band can be perfectly linearly predicted by the variations in the other.  It is clear from Fig.~\ref{fig:lags} that the coherence is high up to frequencies $\gtrsim 10^{-4}$\,Hz.  At frequencies above $\gtrsim 3-4 \times 10^{-4}$\,Hz, the Poisson noise begins to dominate anyway, as shown by the PSD (see Section~\ref{sec:psd}).

The middle panel of Fig.~\ref{fig:lags} shows the phase lags, $\phi$, while the lower panel shows the time delays, $\tau$, both as a function of frequency.  They are simply related by: $\tau = \phi / 2\pi\nu$.  Here, a positive delay signifies that the response of the hard band (i.e. 1--4\,keV) is delayed with respect to variations in the soft band (i.e. 0.3--1\,keV); i.e. a `hard lag'.  A negative delay would imply a `soft lag'.  While at frequencies $> 10^{-4}$\,Hz the time delays are consistent with zero lag, there are some clear deviations at lower frequencies.  While there are no obvious hard delays, the time delays do appear negative in the $4-5 \times 10^{-5}$ and $8-10 \times 10^{-5}$\,Hz frequency ranges, with measured time delays of $\tau = -940 \pm 380$ and $-940 \pm 400$\,s, respectively.  The red data in Fig.~\ref{fig:lags} show the cross-spectral products when including the 2003 and 2013 {\it XMM-Newton} observations in addition to the 2014 data.  It can be observed that the measurements are reasonably consistent between the two cases.

In order to assess the robustness of the lag measurement, we follow the method employed in \citet{DeMarco13} to analyse the incidence of soft lags in a large sample of variable AGN.  Here, we use extensive Monte Carlo simulations to test the reliability of the lag detections against spurious fluctuations (i.e. produced by Poisson / red noise).  Following the method of \citet{TimmerKonig95}, we simulated 1\,000 pairs of stochastic light curves based on fitting the underlying PSD in each of the 0.3--1 and 1--4\,keV bands with a bending power law (i.e. in the same way as described in Section~\ref{sec:psd}).  The simulated light curves were scaled to the mean count rate of the observed light curves and were produced with the same background rate and Poisson noise level as that which we observe in the two selected energy bands.  

We then computed cross-spectral products for each pair of simulated light curves having imposed a lag of zero phase (i.e. $\phi = 0$).  We adopted the same time sampling, segment length, light curve length and frequency-binning factor as we use with the real data.  Then, any observed frequency-dependent time delays in the simulated light curves can be assumed to arise from a statistical fluctuation.  \citet{DeMarco13} test the significance of each lag in their sample by defining a `sliding-frequency' window containing the same number of consecutive frequency bins, $N_{\rm w}$, which contain the observed lag profile (in this instance, $N_{\rm w} = 1$).  They then compute the figure of merit $\chi = \sqrt{\Sigma(\tau/\sigma_{\tau})^{2}}$ at each step over the frequencies below which Poisson noise dominates, recording the maximum value.  Recording the number of times $\chi$ from the simulated data exceeds the value from the real data allows us to estimate the probability of observing such a lag by chance.  In this instance, our observed soft lags are significant at the $> 95$\,per cent level.  As such, this appears to be the first detection of a negative X-ray time delay in Ark\,120.

\begin{figure}
\begin{center}
\rotatebox{0}{\includegraphics[width=8.4cm]{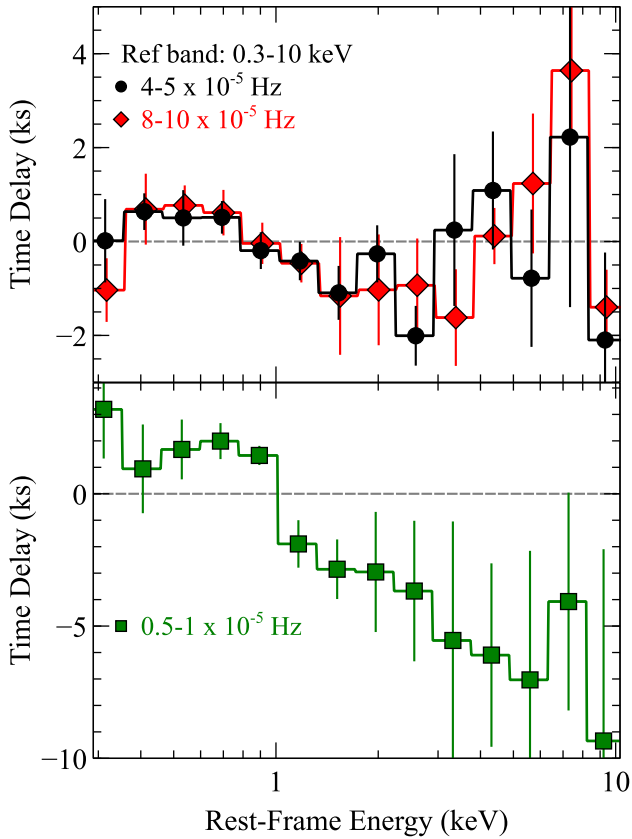}}
\end{center}
\vspace{-15pt}
\caption{EPIC-pn Lag-energy spectra of Ark\,120 computed against a broad reference band (0.3--10\,keV - minus the band of interest) across 14 energy bins with roughly equal logarithmic spacing.  Upper panel: the 2014 data are shown over two frequency ranges: $4-5 \times 10^{-5}$\,Hz (black circles) and $8-10 \times 10^{-5}$\,Hz (red diamonds).  Lower panel: the combined data (2003+2013+2014) are shown over a lower frequency range: $0.5-1 \times 10^{-5}$\,Hz (green squares).}
\label{fig:lag-energy}
\end{figure}

In addition to the frequency-dependence of the lags, they may also show significant energy dependence.  As such, we investigated the energy dependence of the negative lags observed in Fig.~\ref{fig:lags}.  Here, the cross-spectral lag is calculated for a series of consecutive energy bands against a broad reference band, which we keep constant, over a given frequency range (e.g. \citealt{Uttley11}, \citealt{ZoghbiUttleyFabian11}, \citealt{AlstonDoneVaughan14}, \citealt{LobbanAlstonVaughan14}).  Here, we use a reference band covering the full 0.3--10\,keV energy range minus the band of interest\footnote{The choice of reference band has the effect of producing an offset in the resultant spectrum.  For example, computing lag-energy spectra against a 1--4\,keV reference band resulted in lag-energy spectra that were similar in shape to those shown in Fig.~\ref{fig:lag-energy}, just offset on the $y$-axis.}.

In Fig.~\ref{fig:lag-energy}, we show the lag-energy spectra over the $4-5 \times 10^{-5}$ (black) and $8-10 \times 10^{-5}$\,Hz (red) frequency ranges.  Here, a positive lag indicates that the given energy band is delayed with respect to the reference band.  Note that errors on the individual lag estimates in each band were calculated using the standard method of \citet{BendatPiersol10}.  There is a slight hint of roughly log-linear energy dependence at lower energies with the magnitude of the soft lag increasing while moving towards higher energies.  Curiously, both lag energy spectra appear to then show a hint of a peak at higher energies --- in particular, at $\sim$7\,keV (although with large uncertainties).  This appears similar to the Fe\,K lags reported in other sources (e.g. \citealt{Zoghbi12,Kara13,AlstonDoneVaughan14,Kara14}).

Finally, the green curve in Fig.~\ref{fig:lag-energy} then shows the lag-energy spectra derived from cross-spectral products using all six {\it XMM-Newton} observations with a segment length of 110\,ks --- i.e. the longest segment allowed, defined by the length of the observation from 2003.  This allows us to extend the analysis down to the lowest possible frequencies: $\sim$9 $\times 10^{-6}$\,Hz.  Curiously, the lag appears to show strong energy dependence, again with the magnitude of the negative time delay increasing towards higher energies, up to a time delay, $\tau$, of several ks (although with increasingly larger uncertainties).  Given that, compared to other AGN, this frequency appears unusually low for observing a soft lag with such energy dependence combined with the fact that we only have six measurements of the lags at such low frequencies, we emphasise that we must exercise caution in its interpretation.

To summarize, this long {\it XMM-Newton} observation of Ark\,120 has revealed the presence of a significant high-frequency soft-X-ray lag for the first time in this source, with the 0.3--1\,keV emission lagging behind the 1--4\,keV emission with a time delay of $\sim$900\,s.  Modest energy-dependence of the soft lag is observed with the magnitude of the lag increasing towards higher energies in a roughly log-linear fashion, with an additional hint of an Fe\,K lag in the $\sim$6--7\,keV band.  Finally, similar energy dependence of the lag may also be present at the lowest-observable frequencies.

\section{Swift monitoring campaign} \label{sec:swift_monitoring}

Here, we report on the results of our analysis of the {\it Swift} monitoring campaign of Ark\,120 undertaken in 2014/15.

\subsection{Investigating the variability of Ark\,120 with the {\it Swift} XRT} \label{sec:xrt_variability}

\begin{figure*}
\begin{center}
\rotatebox{0}{\includegraphics[width=17.8cm]{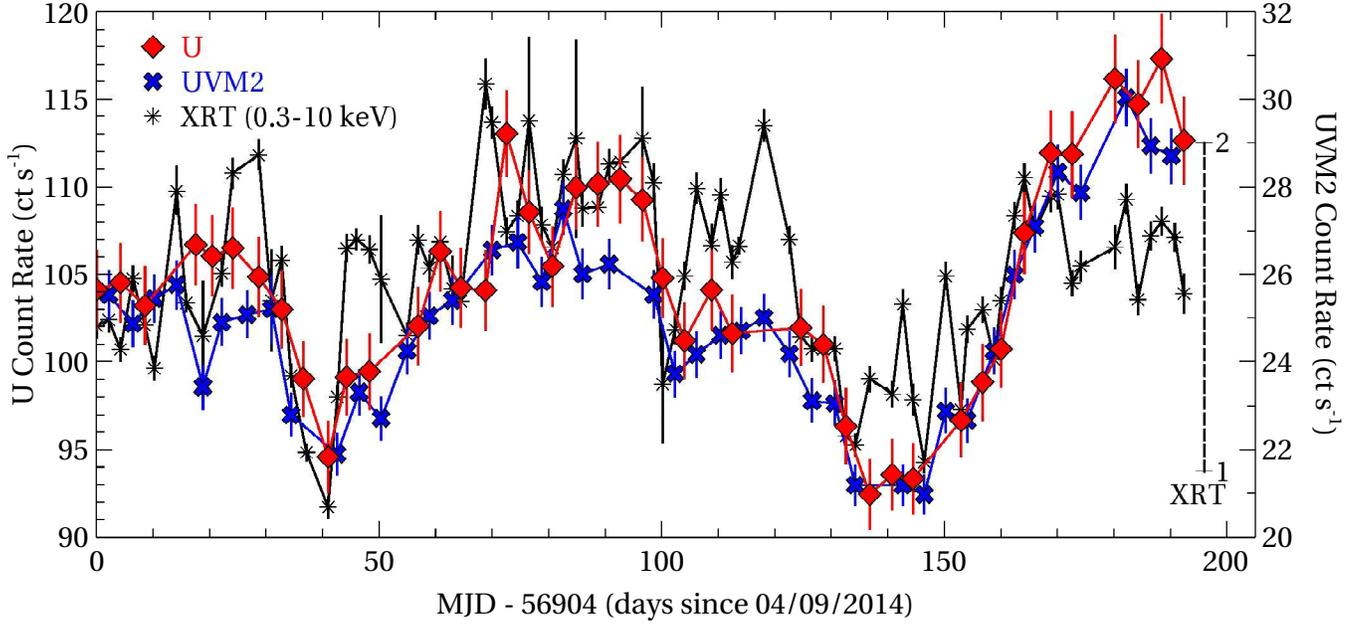}}
\end{center}
\vspace{-15pt}
\caption{The {\it Swift} UVOT light curve of Ark\,120 showing the corrected count rates in the U and UVM2 bands.  Each point corresponds to a single image / observation.  The 0.3--10\,keV XRT light curve is overlaid with an additional y-axis scale for comparison.}
\label{fig:u_uvm2_xrt_lc}
\end{figure*}

In Fig.~\ref{fig:u_uvm2_xrt_lc}, we show the broad-band 0.3--10\,keV {\it Swift} XRT light curve, consisting of the WT-mode corrected count rates obtained from each of the 84 useful observations obtained in 2014/15, covering $\sim$200\,days with an observed sampling rate of $\sim$2 days.  The long-term light curve is observed to be variable, varying peak-to-peak by a factor of $\sim$2, with significant X-ray variability observed on the timescale of a few days.  Superimposed on the XRT light curve are the corresponding U-band and UVM2-band light curves obtained with the UVOT, which we return to in Section~\ref{sec:swift_optical_uv_x-ray}.  

\begin{figure}
\begin{center}
\rotatebox{0}{\includegraphics[width=8.4cm]{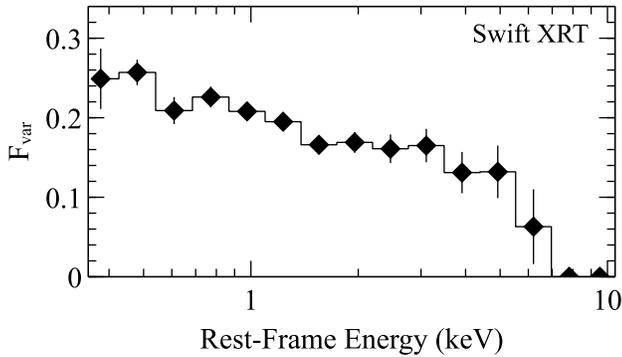}}
\end{center}
\vspace{-15pt}
\caption{The fractional variability of the long-term {\it Swift} XRT light curve in roughly equal logarithmically-spaced bins.}
\label{fig:xrt_fvar}
\end{figure}

In Fig.~\ref{fig:xrt_fvar}, we show the XRT fractional variability, $F_{\rm var}$ (see equation~\ref{eq:f_var}), over a series of energy bins with roughly equal logarithmic spacing.  It can be seen that the long-term {\it Swift} XRT variability gradually increases towards lower energies, peaking at $\sim$30\,per cent.  Note that $F_{\rm var} = 18.7 \pm 0.5$\,per cent over the full broad-band 0.3--10\,keV energy range.  We also investigated the hardness ratio as a function of source flux using the same method described in Section~\ref{sec:xmm_variability} and, again, fitting a simple function of the form: $HR = aCR + b$, finding $a = -0.10$ and $b = 0.38$.  These results are consistent with \citet{Gliozzi17} and confirm modest softer-when-brighter behaviour in this source.  We did also fit the time-average {\it Swift} XRT spectrum and performed additional flux-resolved spectroscopy, finding results consistent with \citet{Gliozzi17}, who perform a detailed spectral analysis.  We refer to that paper for discussion and analysis of the {\it Swift} XRT spectrum.

\subsection{Simultaneous optical/UV/X-ray monitoring with the {\it Swift} UVOT and XRT} \label{sec:swift_optical_uv_x-ray}

As stated in Section~\ref{sec:uvot}, the UVOT on-board {\it Swift} obtained 86 individual observations of Ark\,120, alternating between the U ($\sim$3\,465\AA) and UVM2 ($\sim$2\,246\AA) filters with typical exposure times of $\sim$1\,ks.  The observed sampling rate was $\sim$2\,days (or $\sim$4\,days for a given filter).  In Fig.~\ref{fig:u_uvm2_xrt_lc}, we show the overlaid light curves obtained in the U and UVM2 bands.  The UV data are clearly variable, with the U-band varying by up to $\sim$30\,per cent and the UVM2-band varying by up to $\sim$50\,per cent on the timescale of tens of days.  The amplitude of variability of the two bands, expressed in terms of fractional variability (see equation~\ref{eq:f_var}), is $F_{\rm var} \sim 5.5$ and $F_{\rm var} \sim 8.2$\,per cent, respectively (compared with $F_{\rm var} \sim 16$\,per cent for the X-ray band).

The U and UVM2 bands appear to be closely correlated with a series of coincident peaks and dips in the two corresponding light curves.  Superimposed on the UVOT light curves is the 0.3--10\,keV XRT light curve with an additional y-axis scale for ease of comparison.  Despite displaying more significant short-term variability, the X-rays also appear to track the long-term UV changes well.

\begin{figure}
\begin{center}
\rotatebox{0}{\includegraphics[width=8.4cm]{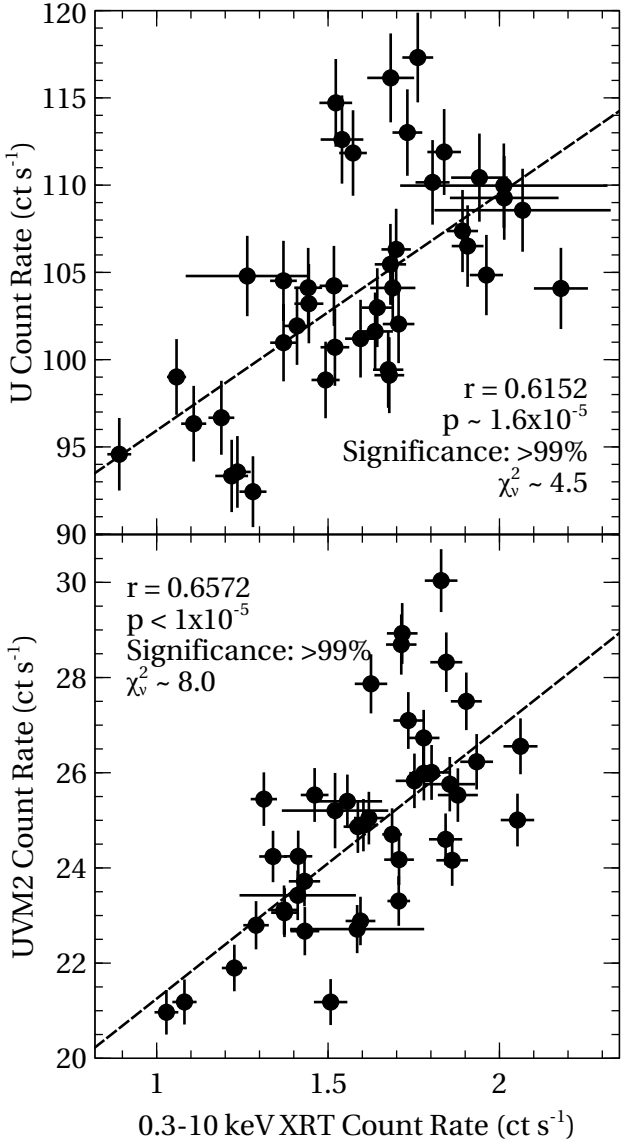}}
\end{center}
\vspace{-15pt}
\caption{The 0.3--10\,keV XRT count rate plotted against the U (upper panel) and UVM2 (lower panel) bands.  In each case, a function of the form $U = aX + b$ is fitted to the data, indicating a clear positive trend.  The reduced $\chi^{2}$ values are included on the plot.}
\label{fig:xrt_vs_uvot}
\end{figure}

In the upper panel of Fig.~\ref{fig:xrt_vs_uvot}, we plot the XRT count rate against the U-band count rate for each of the {\it Swift} observations in which both useful XRT and U-band data were acquired.  The lower panel of Fig.~\ref{fig:xrt_vs_uvot} shows the same but instead for the XRT vs UVM2 data.  Due to the data acquired in a given observation being approximately simultaneous, this is akin to investigating the relationship between the two bands with a time delay, $\tau \approx 0$.

A clear positive correlation is observed in both cases, which we fit with a simple function of the form: $U = aX + b$, where $U$ and $X$ are the count rates in the UVOT and XRT bands, respectively, and $a$ and $b$ are variables.  In both cases, the fit suggests a clear positive trend.  In the XRT vs U-band case, $a = 13.6$ and $b = 82.4$ while in the XRT vs UVM2-band case, $a = 5.7$ and $b = 15.6$.  We also tested the linear relationship between the two bands with the Pearson correlation coefficient, $r$, and find a strong positive correlation in both cases: XRT vs U: $r = 0.6152$ ($p \sim 1.6 \times 10^{-5}$); XRT vs UVM2: $r = 0.6572$ ($p < 10^{-5}$).  In both cases, the positive correlation is significant at the $> 99$\,per cent level.

We also searched for correlations between physical model parameters and the UV flux.  We fitted each of the 86 XRT spectra over the full 0.3--10\,keV band with a simple model of the form: \textsc{tbabs} $\times$ (\textsc{pl} $+$ \textsc{bbody}), where \textsc{pl} is a power law, dominating at hard energies, and \textsc{bbody} is a blackbody component, modelling the soft excess.  Including any additional components in the model would result in it being overly complex given the limited statistics of each individual XRT snapshot, of which there are 43 simultaneous with each of the U and UVM2 observations.

\begin{figure*}
\begin{center}
\rotatebox{0}{\includegraphics[width=17.8cm]{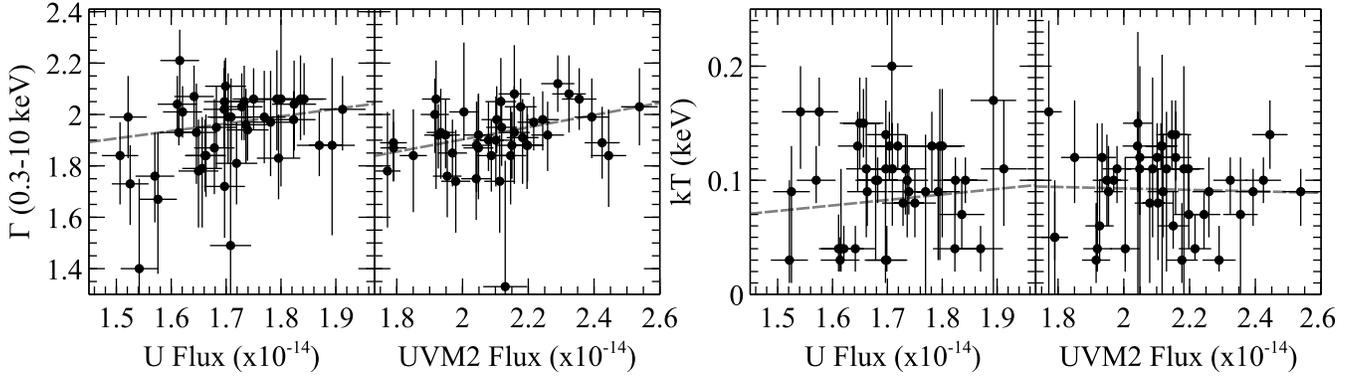}}
\end{center}
\vspace{-15pt}
\caption{The observed fluxes in the {\it Swift} U and UVM2 bands plotted against $\Gamma$ and $kT$ acquired from fits of the form \textsc{tbabs} $\times$ (\textsc{pl} $+$ \textsc{body}) applied to the simultaneously-obtained XRT spectra.  Simple linear trends have been fitted to the plots.  The UVOT fluxes are given in units of erg\,cm$^{-2}$\,s$^{-1}$\,\AA$^{-1}$.  See Section~\ref{sec:swift_optical_uv_x-ray} for details.}
\label{fig:uvot_vs_gamma_kt}
\end{figure*}

In Fig.~\ref{fig:uvot_vs_gamma_kt}, we plot the UVOT fluxes against the power-law slope, $\Gamma$, and the blackbody temperature, $kT$.  We again fit the data with a simple linear model [$\Gamma$($kT$) $= a{\rm U(UVM2)} + b$], finding the following fits (from left to right): $\Gamma = 0.29{\rm U} + 1.48$ ($\chi^{2}/{\rm d.o.f.} = 23.3/41$), $\Gamma = 0.24{\rm UVM2} + 1.43$ ($\chi^{2}/{\rm d.o.f.} = 14.5/41$), $kT = 0.05{\rm U} + 0.00$ ($\chi^{2}/{\rm d.o.f.} = 81.2/41$) and $kT = -0.01{\rm UVM2} + 0.11$ ($\chi^{2}/{\rm d.o.f.} = 42.7/41$).

The linear fits shown in Fig.~\ref{fig:uvot_vs_gamma_kt} hint at the power law slope, $\Gamma$, increasing as the UV flux increases, suggesting that the X-ray slope becomes softer as the source intrinsically brightens.  Similar behaviour is observed in NGC\,7469 (\citealt{Nandra00, Petrucci04}), which the authors discuss in terms of the correlation arising from Comptonization of UV photons to X-ray photons.  However, the correlations only appear to be of moderate strength with values for the Pearson correlation coefficient of $r = 0.360$ ($p = 0.018$) and $r = 0.261$ ($p = 0.091$) for the U vs $\Gamma$ and UVM2 vs $\Gamma$ cases, respectively.  Meanwhile, the UV vs $kT$ correlations are not significant with $r = 0.099$ ($p = 0.528$) and $r = 0.024$ ($p = 0.879$), respectively.

\subsubsection{The cross-correlation function} \label{sec:swift_ccf}

Here, we test for long-timescale correlations between wavelength bands using data from the {\it Swift} monitoring campaign.  The standard tool for measuring correlations between time series is perhaps the cross-correlation function (CCF; \citealt{BoxJenkins76}).  This is a generalized approach to standard linear correlation analysis whereby correlation coefficients are measured between two light curves, allowing for a linear shift in time between the two.  As such, the result is a measure of the degree of correlation, $r$, for a given time shift, $\tau$.  However, the primary requirement for the CCF is that the time series be evenly sampled, which is not the case with the {\it Swift} monitoring campaign.  As such, we estimate the CCF by using the discrete correlation function (DCF; \citealt{EdelsonKrolik88}).  The DCF allows us to calculate the correlation between measured data pairs ($a_{i}$, $b_{j}$), where each given pair has a lag, $\Delta \tau_{ij} = t{j} - t_{i}$.  The complete set of unbinned discrete correlations is first collected (with their associated pairwise lags):

\begin{equation} {\rm UDCF}_{ij} = \frac{(a_{i} - \overline{a})(b_{j} - \overline{b})}{\sqrt{(\sigma^{2}_{a} - e^{2}_{a})(\sigma^{2}_{b} - e^{2}_{b})}}, \label{eq:udcf} \end{equation}

where $e_{a}$ is the measurement error for a given data point in time series, $a$.  This is then binned over the range: $\tau - \Delta\tau/2 \leq \Delta\tau_{ij} < \tau + \Delta\tau/2$, and averaged over the total number of pairs falling within each bin:

\begin{equation} r({\rm DCF}_{ij}) = \frac{1}{M} {\rm UCDF}_{ij}. \end{equation}

Here, $-1 \leq r \leq 1$, where $r = 1$ means that the data are perfectly correlated, while $r = -1$ results from a perfect anticorrelation.  Completely uncorrelated data have $r = 0$.  We note that the mean and variance of each time series in equation~\ref{eq:udcf} may be calculated `globally' or `locally' --- i.e. using the population mean/variance, or, in the latter case, computed using only the data points contributing to that particular lag bin.  We have tried both methods in our subsequent analysis and find no significant difference in our results.

\begin{figure*}
\begin{center}
\rotatebox{0}{\includegraphics[width=17.8cm]{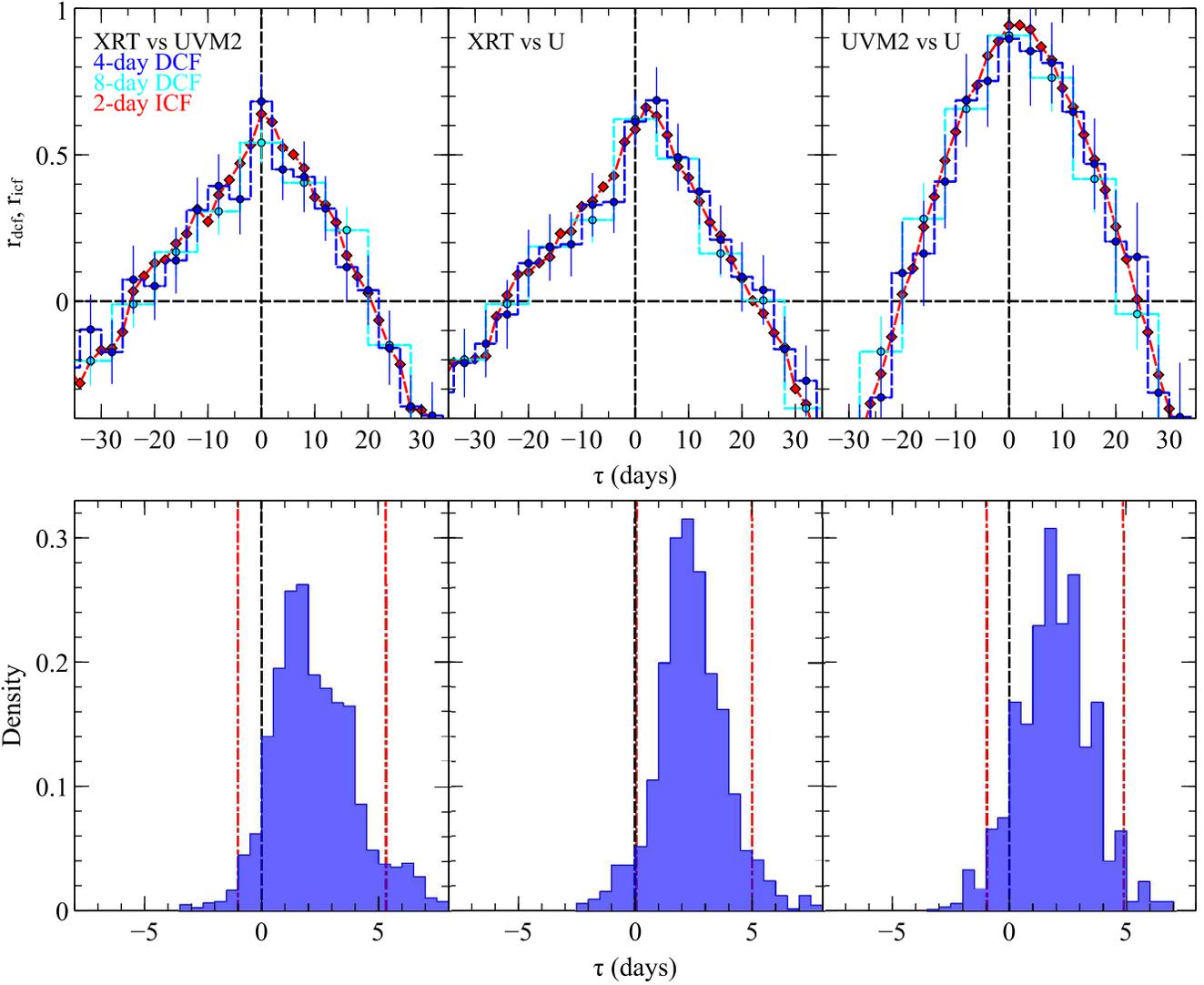}}
\end{center}
\vspace{-15pt}
\caption{Upper panel: the estimated CCF obtained from the $\sim$6-month {\it Swift} monitoring campaign of Ark\,120.  The DCFs are shown with blue and cyan circles with $\Delta \tau = 4$ and $8$\,days, respectively.  The ICF is shown with red diamonds where $\Delta \tau = 2$\,days.  Lower panel: the centroid distributions of the peaks for the ICF analysis.  The red lines correspond to the 90\,per cent confidence intervals based on 10\,000 simulations.  See Section~\ref{sec:swift_ccf} for details.}
\label{fig:ccf}
\end{figure*}

For our CCF estimates, we used the mean count rates from each UVOT exposure and available XRT snapshot, setting the time, $t$, at the centre of each exposure bin.  \citet{Gliozzi17} computed the DCF for Ark\,120 using separate soft and hard X-ray bands while, here, we use the full 0.3--10\,keV band.  While the total length of the {\it Swift} campaign is $\sim$200\,days, at longer lags, fewer pairs contribute to the DCF.  This significantly reduces the certainty on the DCF estimates.  In addition, one should be wary of lags greater than $\sim$1/3-1/2 of the duration of the light curve in the presence of strong ``red-noise'' light curves \citep{Press78}.  Therefore, we computed the DCFs over the range $-35 < \tau < +35$ days using time bin sizes of both $\Delta \tau = 4$ and $\Delta \tau = 8$\,days,\footnote{Note that $\Delta \tau = 4$\,days is roughly equal to the observed sampling rate for the U and UVW2 UVOT filters.} ensuring that there were $> 25$\,ct\,bin$^{-1}$.  We computed these using all three available wavelength bands; i.e. XRT vs U, XRT vs UVM2 and U vs UVM2.  The DCFs are shown in Fig.~\ref{fig:ccf} (upper panel).  A positive lag denotes the `first' wavelength band leading the `second' wavelength band --- in this case, the shorter wavelength leading the longer wavelength emission.

\begin{table*}
\centering
\begin{tabular}{l c c c c c c c}
\toprule
\multirow{2}{*}{Bands} & \multicolumn{2}{c}{4-day DCF} & \multicolumn{2}{c}{8-day DCF} & \multicolumn{3}{c}{2-day ICF} \\
& $r$ (DCF) & $\tau_{\rm lag}$ (days) & $r$ (DCF) & $\tau_{\rm lag}$ (days) & $r$ (ICF) & $\tau_{\rm lag}$ (days) & $\tau_{\rm cent}$ (days) \\
\midrule
X-ray vs UVM2 & $+0.682 \pm 0.095$ & $0 \pm 2$ & $+0.541 \pm 0.071$ & $0 \pm 4$ & $+0.640$ & $0.9 \pm 1.8$ & $0.9 \pm 1.6$ \\
X-ray vs U & $+0.686 \pm 0.113$ & $4 \pm 2$ & $+0.622 \pm 0.068$ & $0 \pm 4$ & $+0.662$ & $2.4 \pm 1.8$ & $2.4 \pm 1.6$ \\
UVM2 vs U & $+0.897 \pm 0.110$ & $0 \pm 2$ & $+0.908 \pm 0.094$ & $0 \pm 4$ & $+0.943$ & $1.7 \pm 2.0$ & $1.7 \pm 1.7$\\
\bottomrule
\end{tabular}
\caption{The CCF estimates for Ark\,120 obtained from $\sim$6 months of monitoring with {\it Swift}.  While the X-ray vs UVM2 and UVM2 vs U CCFs are consistent with zero lag there appears to be a suggestion of the X-rays leading the U-band emission.  Note that the range of $\tau_{\rm lag}$ values for the DCF simply denotes the bin width.  See Section~\ref{sec:swift_ccf} for details.}
\label{tab:ccf}
\end{table*}

The XRT vs UVM2 and UVM2 vs U DCFs are well-correlated at zero lag ($\pm 2$\,days), with $r = +0.682 \pm 0.095$ and $r = +0.897 \pm 0.110$, respectively (both significant at the $>99$\,per cent confidence level), with a clear positive skew in the latter case.  The UVM2 vs U variability correlation is significantly stronger due to the relative smoothness of the light curves.  Meanwhile, the XRT
vs U DCF peaks at $\tau = +4$\,days ($\pm 2$\,days), with $r = +0.686 \pm 0.113$ (again
significant at the $> 99$\,per cent confidence level). We also computed the DCFs with a time-bin size of 8 days (cyan) and found the results to be consistent, although the XRT vs U lag is washed out in the 8-day DCF, peaking at $\tau = 0 \pm 4$\,days.  We also estimate the centroid of the DCFs, $\tau_{\rm cent}$, by taking the mean of all points falling within $0.8 \times \tau_{\rm peak}$, in an approach similar to \citet{Gliozzi17}.  For the XRT vs UVM2, XRT vs U and UVM2 vs U cases, we find $\tau_{\rm cent} \sim 0$, $2$ and $2$\,days, respectively.  The results are tabulated in Table~\ref{tab:ccf}.

In addition to the DCF, we also computed the interpolated correlation function (ICF; see \citealt{GaskellSparke86} and \citealt{WhitePeterson94}).  Here, linear interpolation is performed between data points in first light curve to achieve regular sampling.  The CCF can then be measured for an arbitrary lag, $\tau$, by comparing the interpolated values with the real values from the second light curve.  The reverse process is then done with interpolation applied to the second light curve instead and the results are averaged.  The linear correlation coefficient is then calculated as:

\begin{equation} r({\rm ICF}_{ij}) = \frac{\sum^{N}_{i,j=1} (a_{i} - \overline{a})(b_{j} - \overline{b})}{\left(\sqrt{\sum^{N}_{i=1} (a_{i} - \overline{a})^{2}}\right)\left(\sqrt{\sum^{N}_{j=1} (b_{j} - \overline{b})^{2}}\right)}, \end{equation}

and is repeated for a range of lags in order to find the value of $\tau$ for which the CCF is maximized.  The ICFs were computed with time bin sizes of $\Delta \tau = 2$\,days, which is roughly half of the observed sampling rate in the UVOT light curves.  The ICFs are shown in Fig.~\ref{fig:ccf} (upper panel).  

We estimated uncertainties on the ICF lags using the method of \citet{Peterson98}.  Here, a distribution of lag values is built up by modifying individual light curve points and recalculating the correlation coefficient a number of times, $N$ (in this case, $N = 1\,000$).  We use the combined `FR' (flux randomisation) and `RSS' (random subset selection) methods to modify the light curves.  The FR method randomly deviates each flux point based on a Gaussian distribution of the flux uncertainty (with $\sigma_{a,b} = e_{a,b}$) while the RSS method is similar to ``bootstrapping'' techniques whereby samples are drawn from randomly selected points, reducing the selected sample size by a factor of up to $\sim$1/$e \approx 0.37$ points in order to test the sensitivity of the CCF to individual data points.  

According to the ICF, the peak correlation coefficient comparing the XRT and UVM2 bands is $r_{\rm icf} = +0.640$ with a lag of $\tau = 0.9 \pm 1.8$\,days ($\pm 1.6$\,days centroid), consistent with zero.  However, in the case of the XRT vs U ICF, we find $r_{\rm icf} = +0.662$ peaking at a positive lag of $\tau = 2.4 \pm 1.8$\,days ($\pm 1.6$\,days centroid).  Finally, in the UVM2 vs U case, the ICF has a peak correlation coefficient of $r_{\rm icf} = +0.943$ with a lag of $\tau = 1.7 \pm 2.0$\,days ($\pm 1.7$\,days centroid), again consistent with zero.  See Table~\ref{tab:ccf} for a summary.  In general, this suggests that the shorter wavelength emission may be leading the longer wavelength emission.  Meanwhile, the lower panel of Fig.~\ref{fig:ccf} shows the distribution of the centroid of the peaks in each case, calculated using points with values $r > 0.8r_{\rm peak}$. The vertical dot-dashed red lines show the 90\,per cent confidence intervals based on 10\,000 simulations.

\section{Discussion} \label{sec:discussion}

The purpose of this paper has been to present some of the fundamental timing and variability properties of Ark\,120 through a long $\sim$420\,ks {\it XMM-Newton} campaign plus additional {\it Swift}, {\it RXTE} and {\it NuSTAR} data.  Here, we discuss the results.

\subsection{Spectral decomposition} \label{sec:discussion_spectral_decomposition}

In Section~\ref{sec:spectral_decomposition} we fitted phenomenological models to the broad-band X-ray spectra and analyzed the energy-dependence of the variability.  Both the difference and the rms spectra clearly confirm that the main source of the variability in the 2014 campaign is due to a soft power-law-like component which primarily varies in flux.  Given the relative smoothness of the component and the fact that it can be fitted with a steep power law, this is likely consistent with the cooler Comptonization component detected in paper IV (also see \citealt{Mallick17}).  The soft component appears to dominate the variability on timescales corresponding to frequencies in the range $\nu \sim 0.77-5 \times 10^{-5}$\,Hz (corresponding to timescales in the range: 10--130\,ks) and also between observations on timescales of $\sim$days and $\sim$years.  This behaviour is confirmed by low-frequency covariance spectra presented in Section~\ref{sec:covariance}.  Similar behaviour has recently been observed in the variable Seyfert galaxy PG\,1211+143 \citep{Lobban16}.  Meanwhile, the fractional variability is much less pronounced on short timescales and its energy dependence is much flatter.  As such, the X-ray continuum is likely described by a blend of two components with the soft excess varying slowly and independently of the hard X-ray coronal power law, similar to cases such as NGC\,3227 \citep{ArevaloMarkowitz14}, Ton\,S180 \citep{Edelson02}, Ark\,564 \citep{Turner01} and Mkn\,509 \citep{Mehdipour11}.  If the variable, steep soft excess component arises from Comptonization (as suggested by broad-band modelling in paper IV), a possible scenario might involve intrinsic changes to the corona either in terms of the electron temperature or the optical depth.  For example, the corona may be cooler and/or thinner during time periods where the variable soft component is observed to be weaker.

Through the analysis of difference spectra, we also find evidence for a weak component of variable Fe\,K emission.  This is most prominent in the long-term difference spectra (i.e. between the 2013 and 2014 epochs) in both {\it XMM-Newton} and {\it NuSTAR} data and is observed to peak at $\sim$6.2\,keV.  This energy is moderately red-ward of the core of the line at 6.4\,keV and is like the associated red wing of the Fe line.  This feature in the difference spectra is likely the variable component of emission in the 6.0--6.3\,keV range presented in paper II.  This is observed to vary in strength on timescales of $\sim$1\,year and may arise from a short-lived hotspot on the surface of the accretion disc, perhaps illuminated by magnetic reconnection events, located just a few tens of $r_{\rm g}$ from the central black hole.

\subsection{The power spectrum} \label{sec:discussion_psd}

In Section~\ref{sec:psd} we modelled the PSD of Ark\,120 using all available {\it XMM-Newton} EPIC-pn data.  We fitted the broad-band 0-3--10\,keV PSD (plus PSDs in soft and hard energy bands) with a simple model consisting of a power law plus a constant, the latter of which models Poisson noise which begins to dominate at $\nu \gtrsim 3-4 \times 10^{-4}$\,Hz.  From a joint fit to all six observations we find a mean slope of $\alpha = 1.94^{+0.13}_{-0.12}$, with no clear dependence of the slope on energy, time or flux within the measurement uncertainties.  We find that the slope is consistent with the mean value of $\alpha = 2.01 \pm 0.01$ obtained from a large sample of AGN described in \citet{Gonzalez-MartinVaughan12}.

We also tested for the presence of a break in the PSD by fitting a broken-power-law model (equation~\ref{eq:broken_pl}) to the full {\it XMM-Newton} $+$ {\it Swift} $+$ {\it RXTE} PSD, extending the frequency range down to $\sim$10$^{-7}$\,Hz.  The broadband PSD shape in each energy band is broadly consistent with the X-ray PSD shapes of other Seyfert galaxies with power-law slopes of $\sim$1 below the break, and steeper slopes of $\sim$2--3 above the break.  In the broken-power-law models, the best-fitting break frequencies correspond to break timescales, $T_{\rm br} = 2.9-5.8$\,d --- but, given the large errors, $T_{\rm br}$ is consistent with values spanning 0.54--14.6\,d (the average of each band in log space).  We note that the unbroken-power-law models are not formally rejected at high significance; e.g. rejection probabilities are only 56--79\,per cent due to the poor limited temporal frequency coverage.  Thus, while broken-power-law models provide better fits at all bands, with $L_{\rm BRKN}/L_{\rm UNBR}$ indicating breaks at the 2.3--4.5\,$\sigma$ level, we must conservatively conclude that break detections are tentative, not conclusive.  We find no evidence for evolution of PSD shape or break frequency with photon energy.

The break timescale of the PSD and the mass of the black hole have been found to be approximately linearly correlated from studies of numerous AGN (e.g. \citealt{UttleyMcHardyPapadakis02}; \citealt{Markowitz03}).  A simple scaling relation is provided by \citet{Gonzalez-MartinVaughan12}, which links the observed bend timescale and the mass of the black hole for AGN:

\begin{equation} \label{eq:bend_time-scale_vs_mass} {\rm log}(T_{\rm br}) = A{\rm log}(M_{\rm BH}) + C,  \end{equation}

where $T_{\rm br}$ is the time-scale of the break in days, $M_{\rm BH}$ is the mass of the black hole in units of $10^{6}\,M_{\odot}$ and $A$ and $C$ are coefficients with values of $1.09 \pm 0.21$ and $-1.70 \pm 0.29$, respectively.  For the black hole mass of Ark\,120 ($M_{\rm BH} = 1.5 \times 10^{8}$\,$M_{\odot}$), this predicts bend timescales ranging from $0.85-26.3$\,d, consistent with the observed PSD break measured here.

Additionally, \citet{McHardy06} extend the mass-timescale relation to include an extra dependence on bolometric luminosity, $L_{\rm bol}$.  Here, an additional term is included in equation~\ref{eq:bend_time-scale_vs_mass}: $B{\rm log}(L_{\rm bol})$, where $L_{\rm bol}$ is the bolometric luminosity in units of $10^{44}$\,erg\,s$^{-1}$.  In this instance, the best-fitting values, as derived by \citet{Gonzalez-MartinVaughan12}, are $A = 1.34 \pm 0.36$, $B = -0.24 \pm 0.28$ and $C = -1.88 \pm 0.36$.  For Ark\,120, taking a value of $L_{\rm bol} = 2 \times 10^{45}$\,erg\,s$^{-1}$, this predicts $T_{\rm br} = 5.3$\,d (although with large uncertainties), also consistent with our observations.

\subsection{X-ray time lags} \label{sec:discussion_x-ray_time_lags}

In Section~\ref{sec:x-ray_time_lags}, we presented an analysis of the X-ray variability of Ark\,120 on short timescales through Fourier-based analysis.  We computed frequency-dependent time lags and detected the presence of a soft lag whereby the 0.3--1\,keV band emission lags behind the harder 1--4\,keV band emission.  In Fig.~\ref{fig:lags}, such soft lags appear prominent in the $4-5 \times 10^{-5}$ and $8-10 \times 10^{-5}$\,Hz frequency ranges with measured time delays of $\tau = -940 \pm 380$ and $-940 \pm 400$\,s.  This is the first detection of a soft lag in Ark\,120.

The energy-dependence of the lags are shown in Fig.~\ref{fig:lag-energy} against a constant broad reference band.  There is a hint of a modest log-linear energy dependence with the magnitude of the lag increasing towards higher energies with an additional hint of a peak in the $\sim$6--7\,keV band, similar in shape to the high-frequency Fe\,K lags now observed in a range of AGN (e.g. \citealt{AlstonDoneVaughan14, Kara14, Zoghbi14}).  The Fe\,K lags are often associated with small-scale reverberation from material close to the black hole.  Additional energy-dependence is found with the lowest-frequency lag (when including all six observations) at $\nu \sim 9 \times 10^{-6}$\,Hz.  Curiously, this energy-dependence appears to be much stronger, again with the magnitude of the soft lag increasing towards higher energies, compared to the broad reference band, peaking at a few ks.

A scaling relation between $M_{\rm BH}$ and amplitude/frequency of the soft lag is presented by \citet{DeMarco13} based on a sample of 15 AGN.  The following relations were found for the relationship between observed frequency, $\nu$, and time lag, $|\tau|$, with $M_{\rm BH}$: log\,$\nu = -3.50[\pm0.07] - 0.47[\pm0.09]$\,log\,$(M_{\rm BH})$ and log\,$|\tau| = 1.98[\pm0.08] + 0.59[\pm0.11]$\,log\,$(M_{\rm BH})$, where $M_{\rm BH}$ is the black hole mass in units of $10^{7}$\,$M_{\odot}$.  In the case of Ark\,120 ($M_{\rm BH} = 1.5 \times 10^{8}$\,$M_{\odot}$), this predicts the frequency of the soft lag to lie in the range: $\nu = 5.9 \times 10^{-5} - 1.3 \times 10^{-4}$\,Hz.  This is consistent with the $8-10 \times 10^{-5}$\,Hz lag and marginally consistent with the $4-5 \times 10^{-5}$\,Hz lag.  Meanwhile, the scaling relations predict a time delay in the range: $|\tau| = 290-760$\,s, which is consistent with our measured time delay within the uncertainties.

The most popular model to explain high-frequency soft lags involves reverberation of the primary X-ray emission by material close to the black hole, perhaps via reflection (e.g. \citealt{ZoghbiUttleyFabian11,Fabian13,Uttley14}).  In such a scenario, the observed time lag roughly corresponds to the distance between the primary and reprocessed emission sites.  In the case of Ark\,120, a time delay on the order of $\sim$1\,ks would correspond to a distance of a $\sim$few $r_{\rm g}$.  As such, the scaling of the characteristic timescales compared to lower-mass AGN (e.g. 1H0707$-$495; \citealt{ZoghbiUttleyFabian11}) may be consistent with the disc reverberation scenario.  Additionally, in a number of sources, the energy-dependence of the high-frequency soft lags have shown evidence for features in the Fe\,K band (e.g. see \citealt{Zoghbi12,Kara13,AlstonDoneVaughan14,Kara14}), which have been associated with the small-scale reverberation model.  In Fig.~\ref{fig:lag-energy}, we find there is a hint of a peak in the energy-dependence of the soft lag of Ark\,120 at $\sim$6--7\,keV, which appears similar to the Fe\,K lags detected in other sources.  However, we do note that the inferred distance above is slightly at odds with our Fe\,K modelling in paper II and our detailed broad-band spectral modelling in paper IV, in which we infer that the bulk of the Fe emission arises from $\sim$tens of $r_{\rm g}$ from the black hole.  If the soft excess were to arise purely from reflection with the X-rays originating in a highly compact corona - e.g. the lamppost model geometry - then the discrepancy could, in part, have contributions from systematic errors in either of the inferred distances.  On the other hand, we may consider alternative geometries, such as a more extended X-ray-emitting corona (e.g. see paper II).  Finally, it is worth noting, though, that the estimated size scale of a $\sim$few $r_{\rm g}$ for the reprocessor may actually be significantly larger when considering the effect of dilution from continuum emission in the reprocessing band.  In particular, in paper IV, we find that the soft excess is dominated by Comptonization (largely arising from the fact that the relativistic reflection parameters are unable to be reconciled between the soft excess and the Fe\,K band).  As such, it may be the case that our high-frequency lag measurement does not directly reflect `corona-to-reflected-only' light travel paths.

Finally, we do not detect the low-frequency hard lag, which is now observed to be a common feature of both variable AGN and XRBs.  Low-frequency hard lags are an important phenomenon as they may be ubiquitous in accreting black hole systems with the frequency and amplitude of the lag scaling with black hole mass.  As such, they likely carry important information about the structure of accretion flows surrounding black holes.  Our non-detection of such a lag likely arises from the fact that the black hole mass of Ark\,120 is large ($\sim$1.5 $\times 10^{8}$\,$M_{\odot}$) and so pushes the frequency of the hard lag below our observed frequency range (i.e. $< 7.7  \times 10^{-6}$\,Hz) with our 130\,ks {\it XMM-Newton} observations.

\subsection{Optical/UV monitoring} \label{sec:discussion_optical_uv}

In Sections~\ref{sec:xmm_variability},~\ref{sec:swift_optical_uv_x-ray} and~\ref{sec:swift_ccf}, we analyzed the optical and UV properties of Ark\,120 through {\it XMM-Newton} OM data and a $\sim$6-month monitoring campaign with the UVOT on-board {\it Swift}.  The OM UV data are well correlated with the long-term X-ray variations (see Section~\ref{sec:xmm_variability}) and are observed to track the long-term X-ray flux --- e.g. mimicking the drop in brightness of the source in 2013 --- and varying by up to $\sim$80\,per cent on timescales of $\sim$years.  This huge change in flux may be too large to arise from reprocessing and so, a long timescale suppression of the observed flux, such as that observed in the 2013 epoch, might be due to an intrinsic decrease in the accretion rate of the source.  While the within-observation variability with the OM is far less pronounced, variations of a few per cent can be observed on timescales of $\sim$days.  On these timescales, the shorter-wavelength emission (i.e. UV) displays stronger fractional variability than the longer-wavelength emission (i.e. optical) with a steady decrease in flux over the timescale of the 2014 {\it XMM-Newton} campaign.  The lower variability in the longer wavelength bands (e.g. V, B, U) may be expected due to increased dilution from the host galaxy whereas the shorter wavelength bands are closer to the peak of the UV bump where  more intrinsic variability may be expected.  Meanwhile, an upturn in the UVW2 ($2\,120$\,\AA) flux is observed at the end of the campaign, hinting at a delayed response to the increase in X-ray flux $\sim$2\,days previously.

In Section~\ref{sec:swift_optical_uv_x-ray}, we show the $\sim$6-month UV/X-ray light curves obtained with {\it Swift}.  A clear linear positive correlation is observed between the U and UVM2 bands.  In addition, the X-rays appear to track these variations with a clear positive correlation observed with a time delay, $\tau \approx 0$, significant at the $> 99$\,per cent level.  In Section~\ref{sec:swift_ccf}, we tested for inter-band correlations as a function of time delay by computing DCFs and ICFs.  Our 4-day and 8-day DCFs show strong correlations between all three wavelength bands, generally peaking at $\tau = 0$\,days, consistent with \citet{Gliozzi17}.  However, we do find a non-zero solution for the XRT vs U-band DCF, which peaks at $\tau = 4 \pm 2$\,days, suggestive of the U-band emission lagging behind the X-rays (also see \citealt{Buisson17}).  Additionally, given the slight positive skew in the UVM2 vs U DCF, we do estimate the centroid, $\tau_{\rm cent}$, to be $\sim$2\,days.  We also compute the ICF for each wavelength band, performing Monte Carlo simulations to estimate the error on the peak-correlation lag.  While the XRT vs UVM2 and UVM2 vs U ICFs are found to be consistent with a lag of zero within the uncertainties, we again find evidence of a time lag between the XRT and U bands, with the U-band emission lagging behind the X-rays with a delay of $\tau = 2.4 \pm 1.8$\,days.

Short-timescale correlations between the optical, UV and X-ray bands have been seen in numerous type-1 Seyfert galaxies, such as MR\,2251-178 \citep{Arevalo08}, Mrk\,79 \citep{Breedt09}, NGC\,3783 \citep{Arevalo09}, NGC\,4051 (\citealt{Breedt10}, \citealt{AlstonVaughanUttley13}) and NGC\,5548 (\citealt{McHardy14}, \citealt{Edelson15}).  The observed variations are commonly discussed in the context of reprocessing in the accretion disc.  In this scenario, the disc is illuminated by primary X-rays, which are reprocessed, resulting in the production of UV and optical photons.  The UV/optical photons will then be delayed with respect to the X-rays where the time lag, $\tau$, is defined by the light-crossing time between the emission sites.  Therefore, one may expect large-amplitude X-ray variations to result in time delays that are both smaller in amplitude and wavelength-dependent.  If the disc is optically-thick, the longer-wavelength emission will be delayed with an expected relation: $\tau \propto \lambda^{4/3}$ \citep{CackettHorneWinkler07}.  Now, while the majority of the optical and UV emission could be ``intrinsic'' (for example, arising from internal viscous heating), it is expected to be largely constant over long timescales, whereas an additional portion of observed emission may vary, arising from the reprocessing of X-rays.

For Ark\,120, we can estimate the location of the reprocessing sites assuming a standard accretion disc that is optically-thick.  We assume $\alpha = 0.1$, $H/R = 0.01$, a central compact X-ray source at 6\,$r_{\rm g}$ above the mid-plane, $L/L_{\rm Edd} = 0.1$ (paper IV) and the mass of the black hole to be $1.5 \times 10^{8}$\,$M_{\odot}$.  Then, using equations 3.20 and 3.21 of \citet{Peterson97}, we estimate emission-weighted radii for the {\it Swift} U (3\,465\,\AA) and UVM2 (2\,246\,\AA) bands of 300 and 160\,$r_{\rm g}$, respectively.  These correspond to respective light-crossing times of $\sim$2.0 and $\sim$1.1\,light-days.\footnote{Contrarily, the viscous timescale is on the order of $> 10^{3}$\,years for a geometrically-thin disk and $\sim$days--weeks for a geometrically-thick disc.}  While we do not detect any significant time delay between the XRT and UVM2 bands, we may not expect to detect one given our data's timing resolution.  Meanwhile, our ICF does suggest an XRT-to-U-band peak time delay of $\tau = 2.4 \pm 1.8$\,days.  This is consistent with the predicted $\sim$2.0\,light-day crossing time.  The absolute change in UV / X-ray luminosities observed here is consistent with a reprocessing scenario, whereby the observed variations in X-rays are a factor of $\sim$3 larger than the UV variations but the X-ray luminosity is a factor of $\sim$2--3 times weaker.  Consequently, the large-amplitude changes in the relatively weak X-rays are sufficient enough to power the smaller-amplitude changes in the relatively large UV luminosity.

We note that some recent results have posed challenges for the predictions of the reprocessing scenario --- namely the fact that larger physical separations between emitting regions than expected from standard accretion disc models have been implied by measured time lags (e.g. \citealt{CackettHorneWinkler07}, \citealt{Edelson15}) and microlensing studies (\citealt{Mosquera13}).  This has led some authors (e.g. \citealt{GardnerDone16}) to suggest that the observed optical/UV lags do not arise from the accretion disc itself but instead arise from reprocessing of the far UV emission by optically-thick clouds in the inner regions of the BLR.  While we do not confirm longer-than-expected lags here, to better define the inter-band time lags and offer further tests of the disc reprocessing model, Ark\,120 would likely be an appropriate target for a more comprehensive {\it Swift} monitoring campaign in order to sample the source properties more extensively in time and flux.

\section*{Acknowledgements}

This research has made use of the NASA Astronomical Data System (ADS), the NASA Extragalactic Database (NED) and is based on observations obtained with the {\it XMM-Newton} satellite, an ESA science mission with instruments and contributions directly funded by ESA Member States and the USA (NASA), the NASA/UKSA/ASI mission {\it Swift} and the {\it NuSTAR} mission, a project led by the California Institute of Technology, managed by the Jet Propulsion Laboratory, and funded by NASA.  This work made use of data supplied by the UK {\it Swift} Science Data Centre at the University of Leicester.  AL acknowledges support from the UK STFC under grant ST/M001040/1.  DP acknowledges financial support from the French Programme National Hautes Energies (PNHE) and the EU Seventh Framework Programme under grant agreement number 312789.  JNR acknowledges support from NASA grant NNX15AF12G.  EN acknowledges funding from the European Unions Horizon 2020 research and innovation programme under the Marie Sk\l odowska-Curie grant agreement No. 664931.

\label{lastpage}

\end{document}